%% file: jcap_article_mlcode.tex
\documentclass[a4paper,11pt]{article}

\usepackage{jcappub} 

\usepackage{graphicx}
\usepackage{booktabs}
\usepackage{amsmath}
\usepackage{amssymb}
\usepackage{amsthm}
\usepackage{subfig}
\usepackage{url}
\usepackage{listings}
\usepackage[]{natbib}

\newcommand{\be}{\begin{equation}}
\newcommand{\ee}{\end{equation}}
\newcommand{\ba}{\begin{eqnarray}}
\newcommand{\ea}{\end{eqnarray}}
\newcommand{\brr}{\begin{array}}
\newcommand{\err}{\end{array}}
\newcommand{\bc}{\begin{center}}
\newcommand{\ec}{\end{center}}

\newcommand{\tell}{\tilde{\ell}}

\title{Multiple Lensing of the Cosmic Microwave Background anisotropies}

\author[a]{M. Calabrese,}
\author[b,e]{C. Carbone,}
\author[a,f]{G. Fabbian,}
\author[c,d,e]{M. Baldi,}
\author[a,f]{C. Baccigalupi}

\affiliation[a]{SISSA, Via Bonomea 256, 34136, Trieste (TS), Italy}
\affiliation[b]{I.N.A.F, Osservatorio Astronomico di Brera, Via Bianchi 46, 23807, Merate (MI), Italy}
\affiliation[c]{Dipartimento di Fisica e Astronomia, Alma Mater
  Studiorum Universita' di Bologna, viale Berti Pichat 6/2, I-40127, Bologna (BO), Italy}
\affiliation[d]{I.N.A.F, Osservatorio Astronomico di Bologna, via Ranzani 1, I-40127
Bologna (BO), Italy}
\affiliation[e]{I.N.F.N. - Sezione di Bologna, viale Berti Pichat 6/2,
  I-40127, Bologna (BO), Italy}
\affiliation[f]{INFN, Sezione di Trieste, Via Valerio 2, I-34127 Trieste, Italy}

\emailAdd{mcalabre@sissa.it}
\emailAdd{carmelita.carbone@brera.inaf.it}
\emailAdd{gfabbian@sissa.it}
\emailAdd{marco.baldi5@unibo.it}
\emailAdd{bacci@sissa.it}

\abstract{
We study the gravitational lensing effect on the Cosmic
Microwave Background (CMB) anisotropies performing a ray-tracing of
the primordial CMB photons through intervening large-scale structures
(LSS) distribution predicted by N-Body numerical simulations with a
particular focus on the precise recovery of the lens-induced polarized
counterpart of the source plane.
We apply both a multiple plane ray-tracing and an effective deflection
approach based on the Born approximation to deflect the CMB photons
trajectories through the simulated lightcone. 
We discuss the results obtained with both these methods together with
the impact of LSS
non-linear evolution on the CMB temperature and polarization power
spectra.  We compare our results with semi-analytical approximations implemented in Boltzmann codes like, e.g., CAMB.
We show that, with our current N-body setup, the predicted lensing
power is recovered with good accuracy in a wide range of multipoles
while excess power with respect to semi-analytic prescriptions is observed in the lensing potential
on scales $\ell \gtrsim 3000$. We quantify the impact
of the numerical effects connected to the resolution in the N-Body
simulation together with the resolution and band-limit chosen to
synthesise the CMB source plane. We found these quantities to be
particularly important for the simulation of B-mode polarization power spectrum.
}
 

\begin{document}
\maketitle
\flushbottom

\input{./intro.tex}
\input{./theory.tex}
\input{./algorithm.tex}
\input{./tests.tex}

\input{./results.tex}

\input{./conclusions.tex}

\acknowledgments 

We acknowledge the use of the publicly available Code for Anisotropies
in the Microwave Background (\lstinline!CAMB!), \lstinline!LensPix! and
the Hierarchical Equalised Latitude Pixel spherical pixelization
scheme (\lstinline!HEALPix!).
The L-CoDECs simulations were carried out on the Power6 cluster at the RZG computing centre in Garching and on the SP6 machine at the ``Centro Interuniversitario del Nord-Est per il
Calcolo Elettronico'' (CINECA, Bologna).
The ray-tracing computations have been performed on the IBM Fermi
cluster at CINECA, with CPU time assigned
under several CINECA class-C calls,  and at the National Energy Research Scientific Computing Center (NERSC), which is supported by the Office of Science of the U.S. Department of Energy under Contract No. DE-AC02-05CH11231.
Part of this work was supported by the InDark INFN Grant. CC
acknowledges financial support to the ``INAF Fellowships Programme
2010''
and to the European Research Council through the Darklight Advanced
Research Grant (\# 291521).
MC acknowledges the
financial support through the INAF funds for the Euclid mission. MB is
supported by the  Marie Curie Intra European Fellowship
``SIDUN''  within the 7th Framework  Programme of the European Commission.

\appendix

\section{Measuring angular power spectrum}\label{appendix:ps}
The lensing potential is extracted from N-body simulation through a
binned map-making procedure of particles contained in a sky pixel. For
this reason when we want to extract its underlying power spectrum we
have to correct for the pixel window function of the \lstinline!HEALPix!
grid. Assuming  an azimuthally symmetric patch, as it is the case for
the full sky, the pixel window function is azimuthally symmetric and
can be used to correct the pseudo power spectrum extracted from the
full sky map using a simple spherical harmonic analysis operation.
A continuous field sampled on the \lstinline!HEALPix!
sphere is a smoothed version of the true underlying field due to the finite pixel size, 
i.e. the value of the field in pixel $i$ is given by
\be
\Phi^{pix(i)} = \int {\rm d}^2 \hat{{\bf n}}w^{(i)}(\hat{{\bf n}})\Phi(\hat{{\bf n}}),
\ee
where $w^{(i)}$ is the window function of the $i$-th pixel as is
given by
\be
w^{(i)}(\hat{{\bf n}}) =
\begin{cases}
\Omega_{pix}^{-1}, & \text{inside pixel $i$} \\
0 & \text{elsewhere}.
\end{cases} 
\ee
Expanding the true field $\Phi$ in terms of spherical harmonics
as
\be
\Phi(\hat{{\bf n}}) = \sum_{\ell m} \Phi_{\ell m} Y_{\ell m}(\hat{{\bf n}}),
\ee
we have
\be
\Phi^{pix(i)} = \sum_{\ell m}w_{\ell m}^{(i)} \Phi_{\ell m};
\label{eq::phi_pix}
\ee
where
\be
w^{(i)}_{\ell m} = \int d^2 \hat{{\bf n}} w^{(i)}(\hat{{\bf n}})
Y_{\ell m}(\hat{{\bf n}})
\label{eq::pixel_w}
\ee
is the spherical harmonic transform of the pixel window
function. The computation of these coefficients for each and every
pixel is required for a complete analysis in the \lstinline!HEALPix! scheme; however this calculation
becomes computationally unfeasible, even for a moderate \lstinline!NSIDE!. Therefore, it is
advantageous to ignore the azimuthal variation and rewrite
Equation~(\ref{eq::pixel_w}) as
\be
w^{(i)}_{\ell m} = w^{(i)}_{\ell} Y_{\ell m}(\hat{{\bf n}}_i),
\label{eq::47}
\ee
defining an azimuthally averaged window function as
\be
w^{(i)}_{\ell} = \frac{4 \pi}{(2\ell+1)} \left[ \sum^{\ell}_{m=-\ell} \left
  | w_{\ell m}
  \right |^2 \right]^{1/2}.
\ee
It follows immediately from Equations~(\ref{eq::phi_pix}) and (\ref{eq::47})  that
the estimate of the power spectrum of the pixelated field
is given by
\be
C_{\ell}^{\Phi,pix} = w^2_{\ell} \left \langle \Phi_{\ell m} \Phi^{*}_{\ell m} \right \rangle,
\ee
where the pixel averaged window function is defined as
\be
w_{\ell} = \left( \frac{1}{N_{pix}} \sum^{N_{pix}-1}_{i=0} \left(w^{(i)}_{\ell} \right)^2
\right)^{1/2}.
\ee
This function is available for $\ell < 4\times$\lstinline!NSIDE! in the
\lstinline!HEALPix! distribution. As we divide the computed power spectrum
by the square of the above function, it is possible to correct the effect of the
pixel window; in our case we act directly 
on the spherical harmonics coefficient $\Phi_{\ell m}$ recovered from our spherical maps,
using the actual $w_{\ell}$.

\bibliographystyle{JHEP}
\bibliography{multilens}

\end{document}

%% file: intro.tex

\section{Introduction}
 
The recent measurements of the Planck satellite \cite{PlanckXVI} has unveiled a
Universe well described by a cosmological model known as
$\Lambda$CDM. In this framework, the Universe is expanding 
accelerated by a Dark Energy (DE)
component well described by a cosmological constant $\Lambda$. Cold Dark Matter
(CDM) is responsible for the matter halos around galaxies and galaxy
clusters, while leptons and baryons take a minor part in
the entire cosmic energy density budget.
The robustness of the $\Lambda$CDM model comes primarily from 
advanced and enduring studies on the 
anisotropies in the CMB, combined with
LSS and Type Ia Supernovae observations.
Cosmological perturbations, scalar
(such as density), vector and tensor ones can be
directly observed as primary anisotropies in the CMB sky, imprinted via Compton
scattering at the epoch of recombination. 
A step further in terms of constraining power on cosmological
parameters will be achieved with the high accuracy measurements of the
CMB polarization provided by the upcoming Planck polarization data and current and future high sensitivity ground-based and balloon-borne experiment like, e.g, EBEX \cite{EBEX}, and POLARBEAR \cite{Polarbear}, SPTpol \cite{SPTpol}, ACTpol \cite{ACTpol}, Spider \cite{Spider} and Keck array \cite{keck}\footnote{See \url{http://lambda.gsfc.nasa.gov/links/experimental\_sites.cfm} for a complete list of all the latest missions and upcoming experiments}. These are designed to measure in particular the so-called B-modes of polarization \cite{Zaldarriaga96, Kamionkowski96} which could provide a direct evidence for primordial gravitational waves (tensor perturbation) generated in many inflationary scenarios \cite{Seljak97} if such signal is detected on the degree scale. The BICEP2 collaboration recently announced a robust measurement of a signal compatible with inflationary B-modes \cite{bicep2} and upcoming dataset will be crucial to understand if the signal is genuinely primordial or contaminated by diffuse astrophysical polarised emission \cite{flauger}. \\*
%
In addition to primary
anisotropies previously described, the so-called secondary anisotropies \cite{Aghanim} are
imprinted in the CMB by the interaction of its photons 
with LSS along their
paths from the last scattering surface to the observer. 
One of the most important sources of secondary anisotropies is the
gravitational lensing effect on CMB photons by on growing matter
inhomogeneities which bend and modify the geodesic path of the
light. 
The net-effect of those deviations is a reshuffling of power in the modes of the
primordial power spectrum of the total intensity
 and gradient-like (E-modes) component of the
CMB polarization. Moreover, lensing distorts the primordial polarization
patterns converting E-modes into B-modes pattern, generating power on
the sub-degree scale where we expect the primordial signal to be negligible (for a complete review, see \cite{LewisChallinor06}).
The progress
towards the detection of lensing in the CMB data has been slow, since measurements of the CMB
precise enough to enable a detection of weak lensing were
not available until recently. Moreover, picking out
non-Gaussian signatures - which arise from mode mixing on different
scales induced by lensing - in the measured CMB sky by
itself is extremely difficult, due to confusion from systematics,
foregrounds, and limited angular resolution.
The first robust detections of the lensing effect were
done using temperature data only by ACT \cite{ACT} and SPT \cite{SPT, vanEngelen}, and later
confirmed by Planck with a significance greater than 25$\sigma$. Only
recently, however, the evidence of lensing was detected for the first
time in polarization data by POLARBEAR \cite{pblens, pbbb} and SPTpol
\cite{hanson13}.\\* 
The interest on CMB lensing for cosmological application lies in the possibility of extracting information about the projected large scale structure potential, and thus to
constrain the late time evolution of the Universe and, e.g. the Dark
Energy and massive neutrinos properties \cite{AcquavivaBacci06,
  Smith06, hu-dark,LewisChallinor06}. A step forward to probe DE would be to cross correlate the CMB lensing with the
observations of the actual lenses in LSS surveys as seen by
independent tracers of the matter distribution. This option has
already been exploited to obtained astrophysical and cosmological
information by, e.g., SPT, ACT and, more recently, by POLARBEAR
collaborations  \cite{Bleem12, Sherwin12,act-cfht, pb-cib,hanson13},
but a major improvement is expected in about a decade with the
observations of the ESA-Euclid satellite. 
This will combine arc-second imaging of billions of galaxies with photometric
redshift accuracy corresponding to the percent level, 
between redshifts $0\lesssim z\lesssim 2$
\cite{Laureijs11,Amendola12}. 
To fully exploit the capabilities of these cross correlation studies \cite{pearson-zahn}, accurate prediction for the common observables are crucial. \\*
The most accurate way to obtain predictions for observables of weak-lensing surveys
is to perform ray-tracing through large, high resolution N-Body numerical
simulations to study the full non-linear and hierarchical
growth of cosmic structures. Although this approaches are
computationally very demanding, they allow to check and balance 
the approximations and assumptions made in widely-used semi-analytic
models, adjusting and extending these models if necessary.
Numerous ray-tracing methods have been developed so far in the context
of both strong and weak gravitational lensing. 
Though exact algorithm are available \cite{Killedar11}, they are not
suitable for application targeting observations of large fraction of
the sky for computational reasons. 
A simpler and popular approach consists in using the matter
distribution in the N-body simulations to calculate 
lensing observables by photon ray-tracing along ``unperturbed'', i. e.
undeflected light paths in the so-called Born approximation
(e.g. \cite{Hilbert07a, Couchman99,Carbone09}). 
In particular, \cite{Carbone13}
applied this technique to study a set of N-Body simulation with
different cosmologies and DE dynamics, to investigate the variation of
the lensing pattern with respect to the standard $\Lambda$CDM model. \cite{antolini}, conversely, showed that the
correct integrated matter distribution used to lens incoming CMB
photons in the Born approximation can be properly
reconstructed using standard lensing reconstruction techniques
\cite{hu-okamoto}. 
However, when facing a complex cosmological structure, we must take
into account that each light ray undergoes several
distortions due to matter
inhomogeneities, i.e. approximating the actual path of a photon instead of adopting a single effective deviation from the unperturbed, line-of-sight integral assumed in Born
approximation. The single effective lens must therefore be replaced by
a multiple-lens (ML) approach, where large volumes of matter are
projected onto a series of lens planes \cite{BlandfordNarayan, Jain00, Pace07, Hilbert09,Becker12} so that the
continuous deflection experienced by a light ray is approximated by finite deflections at each
of the planes. 
A ML algorithm full-sky CMB lensing application was sketched in 
\cite{DasBode} who simulated lensed CMB maps in temperature only with
an angular resolution of $0.9'$. However detailed comparison of the
effective Born approximation method and the ML approach was not
discussed and only the results derived in the Born approximation scenario were presented.\\*
%

In this work, we implemented a multiple plane ray-tracing
algorithm to lens CMB temperature and polarization fields combining the aforementioned
work by \cite{DasBode} and using the lightcone reconstruction from a single N-Body simulation of \cite{Fosalba08}.
The final rationale will be to investigate DE effects in different cosmologies at
arc-minutes scales, where are expected to be most noticeable. At these
scales, the Born approximation is likely expected not to trace with
high accuracy local deviations due to
small-scales inhomogeneities, and thus a more precise and realistic method is needed.
Moreover, in order to be successful, we need to be able to control and
discriminate between physical non-linearities of the
N-Body simulations from numerical issues connected to the various approximations in
both the lensing algorithm and the simulation itself. A detailed
analysis of these issues together with their impact on the lensed CMB
observables will thus be presented.
The present paper is intended to be the
first one of a series investigating the response of CMB lensing upon
DE and/or modified gravity cosmologies or neutrino physics, as well as the
feasibility and the constraining power of cross-correlation studies involving Planck and
simulated Euclid data. 
%
%
This paper is organized as follows: in the Section 2, we introduce
the theoretical background and notation used for our lensing
algorithm. In Section 3 we discuss our ray-tracing technique, which is
then tested and evaluated in Section 4. Section 5 shows results for the lensed
temperature and polarization fields of the CMB,
with particular emphasis on differences and similarities between the
Born approximation and the multiple-lens approach. The last Section
draws the conclusions.

%% file: theory.tex

\section{Weak lensing in Cosmology}\label{sec:theory}
\subsection{Gravitational light deflection}
In this Section, we briefly recall the definition of the relevant
quantities concerning the weak lensing effect that will be used in the
rest of the paper. We refer the reader to more specialised reviews for a general
treatment of the weak lensing effect \cite{Bartelmann10,
  LewisChallinor06}. Following the approach in \cite{DasBode}, we assume a coordinate system
based on physical time $t$, two
angular coordinates $\boldsymbol\theta=(\theta_1,\theta_2)$, and the line-of-sight, radial comoving
distance $\eta$ relative to the observer. In a standard Universe with
a weakly perturbed Friedmann-Lemaitre-Robertson-Walker (FLRW) metric,
a light ray approaching a matter density distribution is deviated by an angle 
\be
d \boldsymbol\alpha = -2\nabla_{\perp} \Psi d \eta,
\label{eq:diff_defl}
\ee
where $d \boldsymbol\alpha$ is called deflection angle, $\Psi$ is the Newtonian
potential and $\nabla_{\perp}$ denotes the spatial gradient on a plane
perpendicular to the light propagation direction. The gradient in Eq.~(\ref{eq:diff_defl}) is defined in
the small-angle limit as $\nabla_{\perp} = (\partial / \partial \theta_1, \partial / \partial \theta_2)$
where $(\theta_1, \theta_2)$ describe a coordinate
system orthogonal to the light ray trajectory. 
The transverse shift of the
light ray position at $\eta$ due to a deflection at $\eta'$ can be thus written as
\be
d{\bf x}(\eta) = D_A(\eta - \eta')d\boldsymbol\alpha(\eta'),
\ee
where $D_A(\eta)$ is the comoving angular diameter distance.
%
In weak lensing calculations, an ``effective''
approach is commonly used, where the effect of
deflectors along the entire line of sight is approximated by a projected
potential computed along a
fiducial un-deflected ray (Born approximation).
The final angular position $ \boldsymbol\theta(\eta) = {\bf
  x}(\eta)/D_A(\eta)$
is therefore given by 
\begin{align}
\label{eq:bornlens}
\boldsymbol\theta(\eta) & = \boldsymbol\theta(0) - \frac{2}{D_A(\eta)}\int_0^{\eta}
{\rm d} \eta' D_A(\eta - \eta') \nabla_{\perp}\Psi =\\
\nonumber &= \boldsymbol\theta(0) + \tilde{\boldsymbol\alpha} (\eta),
\end{align}
where $\tilde{\boldsymbol\alpha}$ is the total effective deflection
from the initial angular position of the light ray at the observer position $\boldsymbol\theta(\eta= 0)$. Note that the integral in Eq.~(\ref{eq:bornlens}) is evaluated along the
light ray trajectory and is thus an implicit equation for $\boldsymbol\theta(\eta)$. 
It is convenient to define $\tilde{\boldsymbol\alpha}$ as the gradient of an effective
potential including the contributions to the final deviation of all the structures present between the observer and the background source plane located at a comoving distance $\eta_s$, i.e. ${\bf \tilde{\boldsymbol\alpha}} = -  \nabla_{\perp} \psi^{eff}$, where
\be
\label{eq:born_phi}
\psi^{\rm eff}(\boldsymbol\theta) = \frac{2}{D_A(\eta_s)}
  \int^{\eta_s}_{0} {\rm d} \eta
  \frac{D_A(\eta_s - \eta)}{D_A(\eta)} \Psi(\boldsymbol\theta ,\eta).
\ee
The latter quantity is commonly known as the lensing potential.
If we consider in particular the case of the weak lensing of CMB anisotropies, $\eta_s$  is the comoving distance to the last scattering surface.
 An effective convergence field can be also defined in a similar manner:
\begin{equation}
\kappa(\boldsymbol\theta) \equiv \frac{1}{2} \nabla^2_{\perp}
  \psi^{\rm eff}(\boldsymbol\theta) 
   = \frac{2}{D_A(\eta_s)} \int^{\eta_s}_{0} {\rm d} \eta
  \frac{D_A(\eta_s - \eta)}{D_A(\eta)} \nabla^2_{\perp}\Psi(\boldsymbol\theta ,\eta).
\label{eq:born_kappa}
\end{equation}


\subsection{Discretised formalism}
\label{sec:discr}
The previous set of equations can be discretised by dividing
the interval between the observer and the source
into $N$ concentric shells, each of comoving thickness $\Delta \eta$.
The matter in the $k$-th shell is projected onto a spherical,
two-dimensional sheet which is positioned in the middle of 
the two edges of the matter shell\footnote{The shell index
  $k$ increases as moving away from the source plane.}, at comoving
distance $\eta_k$. To simplify the following calculations, it is common practice to use angular differential
operators defined on the sphere instead of spatial ones, since we will
be working with 2D spherical projections of the matter distribution. 
We thus can rewrite Eq.~(\ref{eq:diff_defl}) in terms
of the angular gradient $\nabla_{\hat{{\bf n}}}$ as 
\be
d \boldsymbol\alpha = -\frac{2}{D_A(\eta)}\nabla_{\hat{n}} \Psi d \eta.
\ee
Note that the versor ${\bf \hat{n}}$ refers to angular coordinates on the
sphere, or $\boldsymbol\theta = \eta {\bf \hat{n}}$. A photon in the $k$-th
shell at $\eta_k$ is deflected - due to the presence of matter - by an
angle $\boldsymbol\alpha^{(k)}$
which can be approximated by
\begin{align}
\label{eq:gradphi}
\boldsymbol\alpha^{(k)}(\hat{{\bf n}}) &= - \frac{2}{D_A(\eta_k)}
\frac{D_A(\eta_s - \eta_k)}{D_A(\eta_s)} \int^{\eta_k+\Delta \eta /2}
_{\eta_k-\Delta \eta /2} \nabla_{\hat{{\bf n}}}
\Psi(\tilde{\eta}\hat{{\bf n}};\tilde{\eta}) {\rm d}
\tilde{\eta}, \\
\nonumber & = - \nabla_{\hat{{\bf n}}} \psi^{(k)}(\hat{{\bf n}}),
\end{align}
where the 2-D gravitational potential on the sphere is defined as
\be
\Phi^{(k)}(\hat{{\bf n}}) = \frac{2}{D_A(\eta_k)} \int^{\eta_k+\Delta \eta /2}
_{\eta_k-\Delta \eta /2} \Psi(\tilde{\eta} \hat{ { \bf n
}};\tilde{\eta}){\rm d}\tilde{\eta},
\ee
and the corresponding contribution to the lensing potential is given by
\be
\psi^{(k)}(\hat{{\bf n}}) = \frac{D_A(\eta_s - \eta_k)}{D_A(\eta_s)}\Phi^{(k)}(\hat{{\bf n}}).
\ee
In the previous equations, the notation $( \eta \hat{{\bf n}}; \eta )$
indicates that the potential is 
evaluated when the
photon is at the position $\eta \hat{{\bf n}}$ on the sky, at the comoving distance $\eta$ from the observer.
We can
relate the gravitational potential to the mass overdensity in the
shell using the Poisson equation
\be
\nabla^2 \Psi= \frac{4 \pi G}{c^2} \frac{\rho - \bar{\rho}}{(1+z)^2},
\ee
where $\bar{\rho}$ is the mean matter density of the Universe at redshift
$z$. As in \cite{ValeWhite03}, we can integrate the above equation along the line
of sight to obtain the two dimensional version of
the Poisson equation for the $k$-th mass shell:
\be
\nabla^2_{\hat{{\bf n}}}\Phi^{(k)}(\hat{{\bf n}}) = \frac{8 \pi G}{c^2}
\frac{D_A(\eta_k)}{(1+z_k)^2} \Delta^{(k)}_\Sigma (\hat{{\bf n}}), 
\label{eq:poiss}
\ee
where the surface mass density is defined as
\be
\Delta^{(k)}_\Sigma (\hat{{\bf n}}) = \int^{\eta_k+\Delta \eta /2}
_{\eta_k-\Delta \eta /2} (\rho - \bar{\rho}){\rm d} \tilde{\eta}.
\label{eq::deltasigma}
\ee
In Eq~(\ref{eq:poiss}) we have dropped the term containing the
derivatives in the radial direction, ignoring thus 
long wavelength fluctuations along the line-of-sight via
the Limber approximation \cite{Jain00}.
However, as argued by \cite{DasBode,Becker12},
this is, at best, an approximation. In particular \cite{Li2010} showed that this assumption neglects
extra terms that become non-zero in presence of realistic
finite width lens plane. 
The problem arises because 
the matter distribution and, thus, the potential itself may 
become discontinuous at the boundaries if periodic conditions are not
enforced. 
Nevertheless, these corrections to the lens-plane approach adopting the  Born approximation which has been used for this work (see Sect.~\ref{sec:algorithm}) confine this problem to the single shells. In fact, partial
derivatives in the transverse plane
commute with the integral evaluated along the whole line of sight, resulting in the cancellation of
line-of-sight modes as required in the Limber approximation of the integral.
 \footnote{Note for example that assuming a flat-sky approximation, unlike what has been done in this work, is a stronger assumption with respect to the Limber approximation and results in a well-known excess of power on large scales as seen, e.g., by \cite{Hu2000}.}

\noindent
We use the following definition for the convergence field $K^{(k)}$ at
the $k$-th shell,
\be
K^{(k)}(\hat{{\bf n}}) = \frac{4 \pi
  G}{c^2}\frac{D_A(\eta_k)}{(1+z_k)^2}\Delta^{(k)}_{\Sigma}(\hat{{\bf
    n}}), 
\label{eq:kappa1}
\ee
to rewrite Eq.~(\ref{eq:poiss}) as
\be
\nabla^2_{\hat{{\bf n}}}\Phi^{(k)}(\hat{{\bf n}}) = 2 K^{(k)}(\hat{{\bf n}}).
\label{eq:poiss_little}
\ee
The lensing potential on the sphere is related to $K$ via Eq.~(\ref{eq:poiss_little}), and
it can be easily computed by expanding both sides of the Poisson
equation in spherical harmonics, obtaining
the following algebraic relation between the harmonic coefficients of
the two fields:
\be
\Phi_{\ell m} = \frac{2}{\ell(\ell+1)}K_{\ell m}.
\label{eq:kappa_to_phi}
\ee
The monopole term in the lensing potential does not contribute to the
deflection field: therefore to avoid any divergence in the above
equation we can safely set to zero $\Phi_{\ell m}$ for $\ell = 0$ in all calculations.
The quantity $K$ is directy computed when the matter distribution in
the shell is radially projected onto the spherical map; 
as discussed in Sect.~\ref{sec:maps2shell}, it is useful to define an angular surface mass density
$\Delta^{\theta}_{\Sigma}(\hat{{\bf n}})$ as the mass per steradians,
\be
\Delta^{\theta (k)}_{\Sigma}(\hat{{\bf n}}) = \int^{\eta_k+\Delta \eta /2}
_{\eta_k-\Delta \eta /2} (\rho -
\bar{\rho})\frac{D_A(\tilde{\eta})^2}{(1+\tilde{z})^3} {\rm d} \tilde{\eta}.
\ee
such that Eq.~(\ref{eq:kappa1}) can be rewritten as:
\be
K^{(k)}(\hat{{\bf n}}) = \frac{4 \pi
  G}{c^2}\frac{1+z_k}{D_A(\eta_k)}\Delta^{\theta (k)}_{\Sigma }(\hat{{\bf n}}).
\label{eq:key}
\ee
Finally, the vector field $\boldsymbol\alpha(\hat{{\bf n}})$ will be synthesised, 
as described in \cite{Zaldarriaga96,Kamionkowski96}, from the spherical harmonic components of the potential in
terms of spin-1 spherical harmonics.
The multiple-plane lens formalism can be also applied to exploit the
effective or single-plane approximation to lens the CMB.
Eqs.~(\ref{eq:born_phi}) and (\ref{eq:born_kappa}) can be discretised into the following sums,
\be
\label{eq:born_phi_dis}
\psi^{\rm eff}(\hat{{\bf n}}) = \sum_j \frac{D_A(\eta_s -
  \eta_j)}{D_A(\eta_s)} \Phi^{(j)},
\ee
\be
\kappa(\hat{{\bf n}}) = \sum_j \frac{D_A(\eta_s -
  \eta_j)}{D_A(\eta_s)} K^{(j)},
\ee
where we used the previous definitions of quantities on the $j$-th lens.
In the same framework, the convergence $\kappa$ can be seen as
just a weighted projected surface density \cite{Fosalba08,Hilbert09}:
\be
\label{eq:kappadelta}
\kappa(\boldsymbol\theta) = \frac{3 H^2_0 \Omega_{m,0}}{2 c^2} \int {\rm
  d}\eta \:
\delta(\eta, \boldsymbol\theta) \frac{D_A(\eta_s - \eta)
  D_A(\eta)}{D_A(\eta_s) a(\eta)},
\ee
where $\delta$ is the 3D matter density at radial distance $\eta$ and
angular position $\boldsymbol\theta$, $D_A(\eta_s)$ is the position
of the lensing source at the last scattering surface and $a(\eta)$ is
the scale factor at $\eta$.
Based on the definition in Eq.~(\ref{eq:kappadelta}), the
angular power spectrum of the convergence becomes
\be
\label{eq:clkappa}
C^{\kappa\kappa}_{\ell} = \frac{9 H^4_0 \Omega^2_{m,0}}{4 c^4}
  \int^{\eta_s}_0 {\rm d}\eta
  P(k,z)\left(\frac{D_A(\eta_s - \eta)}{D_A(\eta_s) a}\right)^2,
\ee
where $P(k,z)$ is 3D matter power spectrum, computed via the Limber
approximation at $k= \ell / D_A(\eta(z))$, valid for $\ell > 10$ within a few
percent accuracy \cite{ValeWhite03}. The discretised equation reads:
\be
\label{eq:clkappa_dis}
C_{\ell}^{\kappa\kappa} = \frac{9 H^4_0 \Omega^2_{m,0}}{4 c^4} \sum_k
  \Delta \eta \:
  P(\ell/D_A(\eta_k),z_k) \left( \frac{D_A(\eta_s - \eta_k)}{D_A(\eta_s)
    a_k} \right)^2,
\ee
summing all over the $k$ lens plane. Note that the convergence field can be converted into
lensing potential using the Poisson equation, or in terms of the angular
power spectrum:
\be
\label{eq:kappa2lens}
C^{\psi\psi}_{\ell} = \frac{4}{\ell^2(\ell+1)^2}C^{\kappa\kappa}_{\ell}.
\ee

%% file: algorithm.tex

\section{The Algorithm}
\label{sec:algorithm}
In the previous Section we have described the basic formalism and equations to evaluate the weak lensing effects on the full sky. In this Section we proceed outlining the basic steps of the algorithm used to lens the CMB photons throughout:
\begin{itemize}
\item[(i)] starting from an N-Body simulation, we create 3D
  simulated matter distribution around a chosen observer;
\item[(ii)] taking advantage of the proper sampling in redshift of the
  simulation, we select different shells of matter at different times to
  reconstruct our past lightcone and mimic cosmological evolution;
\item[(iii)] we project all the matter in a given shell over a single 2D
  spherical map which acts as lensing plane;
\item[(iv)] we solve the full-sky Poisson equation in the harmonic
  domain and compute the lensing potential map for the single lens plane
  and for the integrated potential of Eq.~\ref{eq:born_phi_dis};
\item[(v)] we use this lensing potential map to lens the
  CMB source plane; 
\item[(vi)] we repeat step (iii)-(v) for all the selected shells, thus
  following the evolution of the source plane
\end{itemize}
\noindent
In our analysis we have used a N-Body simulation of cosmic
structure formation in a flat $\Lambda$CDM universe
with an underlying cosmology described by the following set of cosmological parameters: 
\begin{equation*} 
\{ \Omega_{dm},  \Omega_{b}, \Omega_{\Lambda}, n_s, \sigma_8, H_0
  \} = \{ 0.226, 0.0451, 0.729, 0.966, 0.809, 70.3 \: \rm{Km / s /
    Mpc} \}. 
\end{equation*}
The simulation follows the evolution of the matter distribution in a
 cubic (comoving) volume $($1000 $h^{-1} \text{Mpc})^3$ from redshift
 $z = 10$ to present time using a modified TreePM version of
 \lstinline!GADGET!\footnote{\url{http://www.mpa-garching.mpg.de/gadget}},
 specifically developed to include all the additional physical effects
 that characterize different dark energy models (see \cite{Baldi12} for a
 detailed description of the code). The whole numerical project goes
 under the name of COupled Dark Energy Cosmological Simulation, or
 CoDECS\footnote{\url{www.marcobaldi.it/CoDECS}}. At present, they
 include two distinct set of publicly available runs, the L-CoDECS and the H-CoDECS. 
The L-CoDECS simulations consist in $1024^3$ CDM and $1024^3$ baryon
particles, both treated with collisionless dynamics only, which means that
baryonic particles are not considered as gas particles but just as a
different family of collisionless particles distinguished from CDM. This is done in order to account for
the effect of the uncoupled baryon fraction in cDE models which would not be correctly represented
by CDM-only simulations. The mass resolution at $z = 0$ for this set
of simulations is $M_{CDM} = 5.84 \times 10^{10} M_{\odot} / h$ for CDM and
$M_b = 1.17 \times 10^{10} M_{\odot} /h$ for baryons, while the gravitational softening
is set to $\epsilon_s = 20$ comoving $kpc/h$, corresponding to 0.04 times the mean linear inter-particle
separation. The H-CoDECS simulations are instead adiabatic
hydrodynamical simulations on much smaller scales, which we do not consider in the present work.
In this paper we will analyse only the $\Lambda$CDM simulation of the L-CoDECS suite, while the analysis on different DE models will be discussed in a future paper. 

\subsection{Constructing mass shells}
\label{sec:stacking}
N-Body simulations are usually stored as a series of snapshots, each
representing the simulation box at a certain stage of its evolution. 
As a first step, we fix the observer. We consider the last
snapshot, at redshift $z=0$, and compute the center of mass of the whole simulation. 
This centre represents the origin of our new system
of reference, which sees all the CMB
radiation around it. As we explore the universe around the new origin, the further
we move in space, the more we look back in time and see how structures
develop and grow, until we reach a volume large enough to carry out
the integration for CMB lensing. 
One of the difficulties in this approach is that, even though the size of the simulation box is limited, we need to use the box to reconstruct the full backwards lightcone. Therefore,
we need to replicate the box volume several times in space, so that the
entire observable volume is covered. 
In particular, as described in \cite{Carbone08}, the simulation volume needs to
be repeated  along both the positive and
negative directions of the three principal Cartesian axes x, y,
and z, keeping the origin centered on the observer. 

To construct the all-sky past light cone we exploit the simulation
outputs at different times which are equally spaced in the logarithm
of the scale factor, $log_{10}(a)$, corresponding to an average spacing of
150 Mpc$h^{-1}$ comoving.
The need to repeat the simulation volume due to its finite size
immediately means that, without augmenting large-scale structures, the
maps will suffer from a deficit of lensing power on large angular
scales, due to the finite box size. More importantly, a scheme is
required to avoid the repetition of the same structures along the line
of sight. Previous studies that constructed simulated lightcone maps
for small patches of the sky typically simply randomized each
of the repeated boxes along the past lightcone by applying independent random translations and reflections 
(e.g. \cite{Springel01}). However, in the present application
this procedure would produce artifacts like ripples in the
simulated deflection-angle field, because the gravitational
field would become discontinuous at box boundaries, leading
to jumps in the deflection angle. It is therefore mandatory
that the simulated lensing potential of our all sky maps is
everywhere continuous on the sky, which requires that the
3D tessellation of the peculiar gravitational potential is
continuous transverse to every line of sight.

Following \cite{Carbone08, Carbone09}, our solution is to divide up the volume out to the redshift covered
by the simulation $z_{max}$ into larger spherical shells, each of
thickness 1 Gpc$h^{-1}$ comoving (as the box size)
All the simulation boxes falling into the same larger shell are made to undergo
the same, coherent randomization process, i.e. they are all
translated and rotated with the same random vectors generating a
homogeneous coordinate transformation throughout the shell. But this
randomization changes from shell to shell.
\noindent
See Figure~1 of \cite{Carbone08} for a schematic sketch of this stacking technique.
As already mentioned, the need to repeat the simulation volume due to its finite
size implies that the maps will
suffer from a non proper description of the large angular scales. We
note however that if the box size is sufficiently big like in
e.g. \cite{Watson13} this procedure is no longer necessary, at least
up to the redshift covered by the simulation size.
The final results of the whole process is a series of concentric shells that substitutes our
snapshots. For our specific input N-body simulation, we get 25 matter
shells, building a lightcone spanning from $z=0$ to $z_{max} = 10$.

\subsection{From shells to maps}
\label{sec:maps2shell}
Following the scheme proposed in \cite{DasBode}, we convert the position of a
particle distributed within a 3D matter shell into its angular position on the 2D
spherical map of the (projected) surface matter density. Note
  that among all the particles in the simulation, only the ones falling
  within the radii of the spherical shell of width 150 Mpc$h^{-1}$ are projected onto the 2D
  spherical map. 
We then assign each particle to a specific sky pixel in the \lstinline!HEALPix! \footnote{\url{http://healpix.sourceforge.net}} pixelisation scheme starting from its spherical
coordinates ($\theta$,$\phi$)  and using the \lstinline!ang2pix!
routine of the \lstinline!HEALPix! suite. We place the particle
mass into the pixel so that each sky
pixel reads $\Sigma^{\theta}_p = m_p / \Omega_{pix}$, where
$\Omega_{pix}$ is the area of a pixel in steradians. If $n$
particles fall inside the beam defined by a pixel, the pixel will have a surface mass density of $n\Sigma^{\theta}_p$. In
\lstinline!HEALPix!, the resolution is controlled by the parameter
\lstinline!NSIDE! which determines the number $N_{pix}$ of pixels of equal area
into which the entire sphere is divided through the relation
$N_{pix} = 12 \times $\lstinline!NSIDE!$^2$, so that each pixel has
an area of $\Omega_{pix} = 4 \pi/N_{pix}\: \textrm{sterad}$.
The angular resolution is often expressed through the
number $\theta_{res} = \sqrt{\Omega_{pix}}$.
For a value of \lstinline!NSIDE! set to 2048 (4096), the corresponding
an angular resolution is $1^{\prime}.717$ ($0^{\prime}.858$). \\*
The real interesting quantity in our lensing calculation, in addition to
the surface mass density, is the convergence map $K^{(k)}({\bf
  \hat{n}})$ on the lens-plane $k$. To get this quantity we first
  compute the overdensity maps ($\Delta \Sigma^{\theta}$) using the average surface
mass density of the 2D map. Then we multiply by this map by its geometrical weight $(1+z_k)/D_A(\eta_k)$,
depending on the lens plane distance from the observer and its
redshift, assumed to be an average between the time at the beginning
and the end of the shell (see Eq.~\ref{eq:key}). As a final step, we produce a convergence map
from each shell of the  lightcone which will become our
lensing planes to lens the CMB signal.\\*
%
%
From the convergence map $K^{(k)}(\hat{{\bf n}})$ we then extract the
gravitational potential $\psi^{(k)}$ following
Eq.~(\ref{eq:kappa_to_phi}), using the \lstinline!HEALPix! spherical
harmonics transform (SHT) routines to decompose $K^{(k)}(\hat{{\bf
    n}})$ into its harmonic coefficients $K_{\ell m}$. Note that we
correct the smoothing of the true underlying continuous field on the
pixel scale directly in the harmonic domain when we solve for the
Poisson equation (see Appendix \ref{appendix:ps} for more details).

\subsection{Lensing the CMB}
To propagate the CMB photons through the different shells we adopt a
pixel-based approach first presented in \cite{Lewis05}. Starting from
the $\psi_{\ell m}$ coefficients, we compute the deflection field for
each shell $\boldsymbol\alpha^{(k)}$ evaluating Eq.~(\ref{eq:gradphi})
in the harmonic domain. Being the deflection field purely a gradient
field (i.e. a spin-1 curl-free vector field), it can can be easily evaluated
with a spin-1 SHT. The E and B decomposition of the field reads:
\be
_{1}\alpha^{E}_{\ell m}{}^{(k)}=\sqrt{\ell(\ell+1)}\psi^{(k)}_{\ell m} \qquad _{1}\alpha^{B}_{\ell m}{}^{(k)}=0.
\ee
Once the deflection field is give, each pixel-based method remaps the CMB field
as a function of the position on the sky assuming the lensed signal
observed along a direction ${\bf \hat{n}}$ equal to the signal
coming from another direction ${\bf \hat{n}^{\prime}}$, 
\be
{\bf \hat{n}^{\prime}}{}^{(k)} = \hat{{\bf n}}^{(k-1)} + \boldsymbol\alpha^{(k)},
\ee
where $\hat{{\bf n}}^{(k-1)}$ represents the unlensed position of the
CMB photons at the previous step. To our level of approximation $\| \nabla \Phi \|$ is
assumed to be constant between ${\bf \hat{n}}^{(k-1)}$ and ${\bf
  \hat{n}^{\prime}}{}^{(k)}$, consistent with working out the lensing
potential in the Born approximation between two consecutive
shells. 
In this work we adopted the publicly available code
\lstinline!LensPix! to propagate the CMB signal through all the
lensing shells. \lstinline!LensPix! implements a pixel-based lensing
method using a bi-cubic polynomial interpolation scheme to evaluate the
source plane along the displaced direction. This method has been shown
to be accurate at the sub-percent level to produce temperature and
E-modes signal. However, the recovery of the B-modes of the CMB polarization
is more difficult because B-modes are more sensitive to numerical
effects like the involved resolution and the choice
of the band-limit (i.e. the power cut-off $\ell_{max}$) in the calculation. We will
discuss the impact on the relevant numerical effects in Section~\ref{sect:numerical-pixel-based}
and we refer the reader to \cite{Fabbian13} for a complete discussion
of the numerical problems and accuracy of pixel based methods.\\*
%
%
Finally, note that the simulated lightcone recovers the distribution of
matter in the Universe up to
$z_{max}=10$, and therefore the primordial CMB fields are lensed by LSS
only in this specific redshift interval. In other words, photons are ray-traced
in a Universe evolving from $z_{max}$ to $z=0$. The impact of
high-redshifts contributions is besides the goal of this algorithm,
which will be tested against analytical and semi-analytical
computations which we have modified accordingly to perform CMB lensing
only in this redshift range. 

%% file: tests.tex
\section{Test and Convergence}
\label{sec:test}

\subsection{Sanity Checks}
 
\begin{figure}[!htb]
\centering
\includegraphics[scale=0.6,angle=-90]{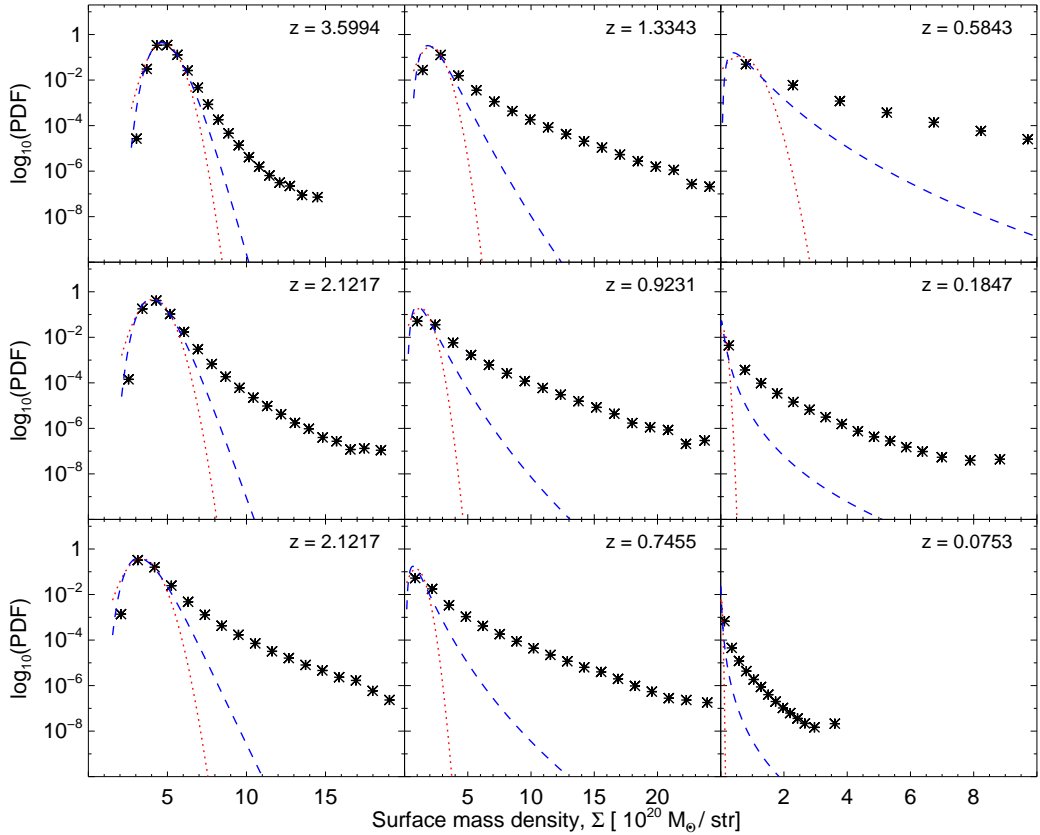}
\caption{
The probability density function (PDF) of the surface
mass density in the lensing-planes (crosses) compared with the
Gaussian (red, dotted line) and the log-normal
(blue, dashed line).
 }\label{fig:PDF}
\end{figure}

In this section we assess the reliability of our code by performing
 sanity checks similarly to \cite{DasBode} to ensure
that all the steps of the algorithm give stable and robust results. 
For the first test, we show that the total mass selected in each 3D
matter slice is equal to the theoretical
mass expected from the assumed cosmological model in the simulation, given by
\begin{equation}
M_{slice}^{theory} = 4 \pi \Omega_{m,0}  \rho_c \bar{\eta}^2 \Delta
\eta, 
\end{equation}
where $\Delta \eta$ is the comoving thickness of the slice at a comoving
(average) distance $\eta$, $\rho_c$ is the critical density and
$\Omega_{m,0}$ corresponds to the present
value of the matter density parameter. We compare this quantity with
the total mass obtained from the surface density maps
($\Sigma^{\theta}$) drawn with our procedure,
\begin{equation}
M_{slice} = \sum^{N_{pix}}_{p=1} \Sigma^{\theta}_p \: \Omega_{pix}, 
\end{equation}
by summing on all pixels of the spherical map.
\begin{figure}
\centering
\includegraphics[scale=0.5,angle=-90]{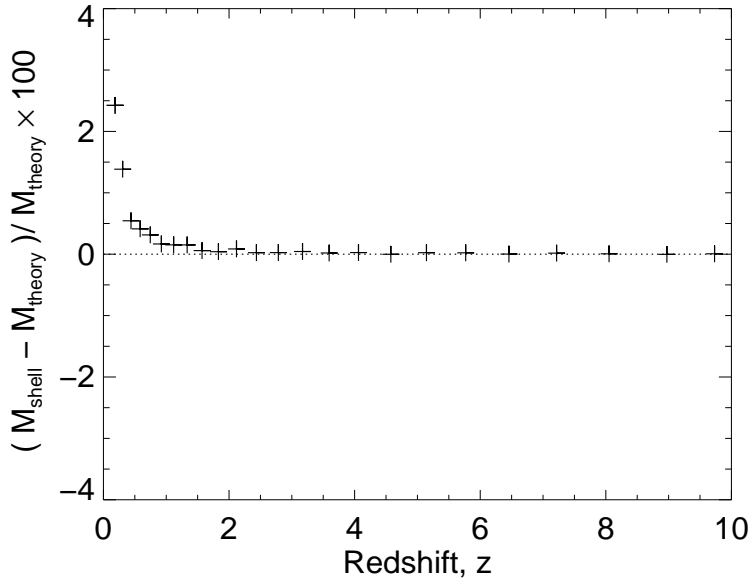}
\caption{
The total mass for each shell compared with the one expected from
theory (fractional difference) as function of redshift.
 }\label{fig::mass_diff}
\end{figure}
Figure~\ref{fig::mass_diff} shows the fractional difference between the
two masses for the different redshifts at which each
spherical map is located. The agreement is very good within a few percents. As similarly found
by \cite{DasBode}, fluctuations respect to the theory appear at low $z$,
due to the tension between the small comoving volume as seen by the
observer, and a highly-clustered matter distribution at late times. Including or excluding large
dark matter halos in the selection process therefore leads to
differences between the mass extracted from the maps and the theoretical one. \\*
As a second test, we make sure that the projection from the simulation box onto the map
has been properly performed. Figure~\ref{fig:PDF} displays the probability density function
(PDF) as recovered from the surface mass density maps, compared
with analytic PDFs, drawn from the data, such as the Gaussian and
log-normal ones (as in \cite{Kayo01,Taruya02}).
The extracted PDFs are quite similar to the ones found by \cite{DasBode},
even if - as already observed by the same authors - the analytical PDFs could not fit well
the data especially at high surface mass density where the
non-linearities becomes important and  where accurate models are yet
unknown.

\subsection{Lensing potential maps}
\label{sec:lensing_pot}
\begin{figure}[!htb]
\centering
\includegraphics[scale=0.6,angle=90]{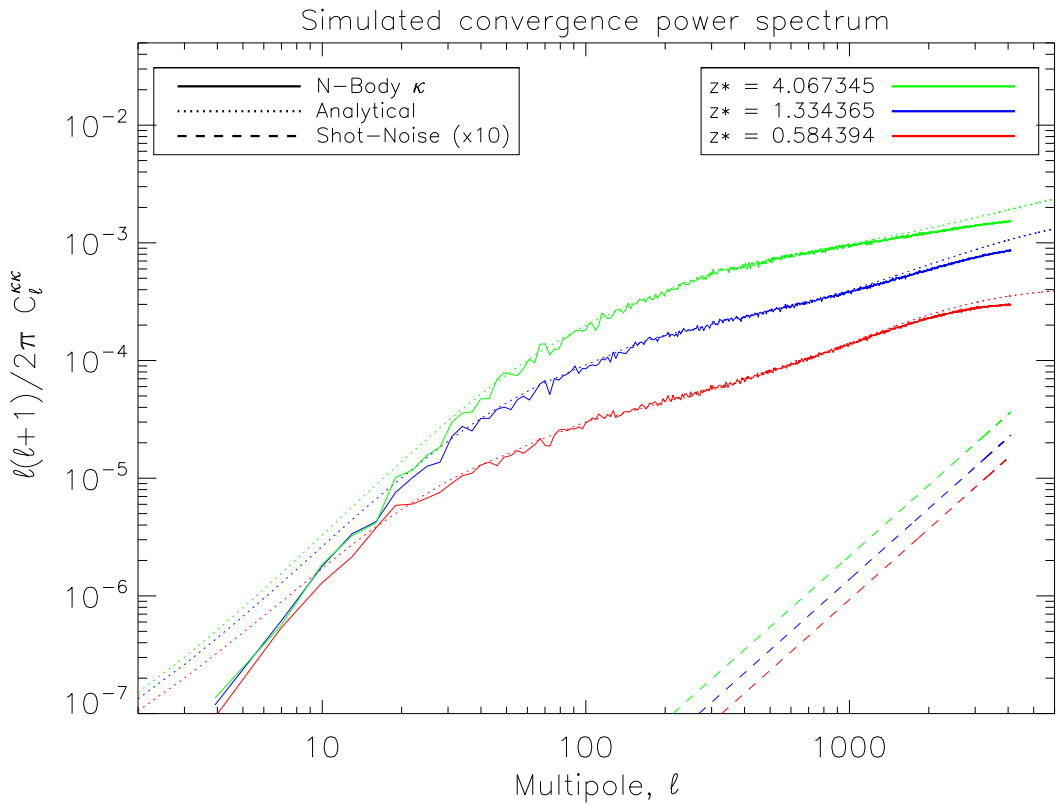}
\caption{
Angular power spectrum from the ``effective'' convergence map
(solid lines) 
compared with theory predictions (dotted line) at three
different redshifts, $z^{*} \simeq 0.6$, 1.3, 4. Dashed lines quantify
the shot-noise contribution for each maps. Note that the shot-noise
power spectra is multiplied by a factor of 10 for visualization
purposes. 
}
\label{fig:born_kappa}
\end{figure}

\begin{figure}[!htb]
\centering
\includegraphics[scale=0.6,angle=90]{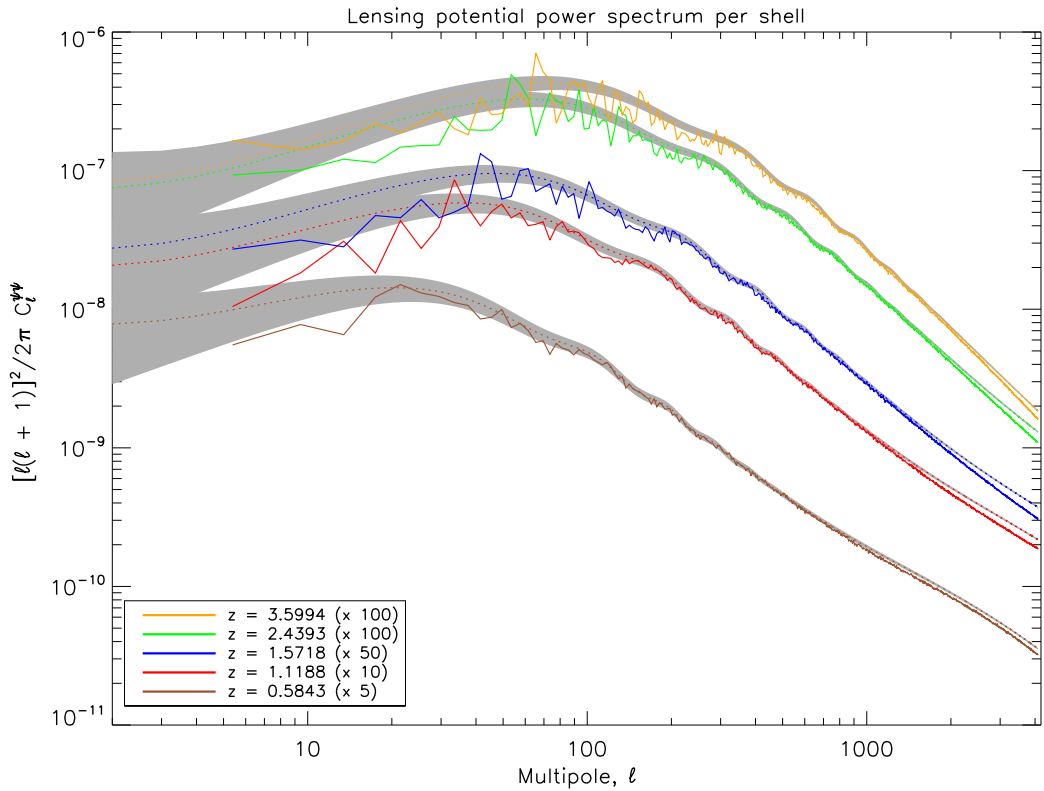}
\caption{
Comparison between the angular power spectrum of the lensing potential computed with
our algorithm (solid lines) and the
analytical results obtained using the Limber equation (dotted lines) at different redshift. The
spectra have been multiplied by a constant factor for displaying
purposes. The grey area displays the cosmic variance $1\sigma$
uncertainty for the theoretical spectra.}
\label{fig:phi_camb}
\end{figure}

\begin{figure}
\centering
\includegraphics[width=.58\textwidth,angle=90]{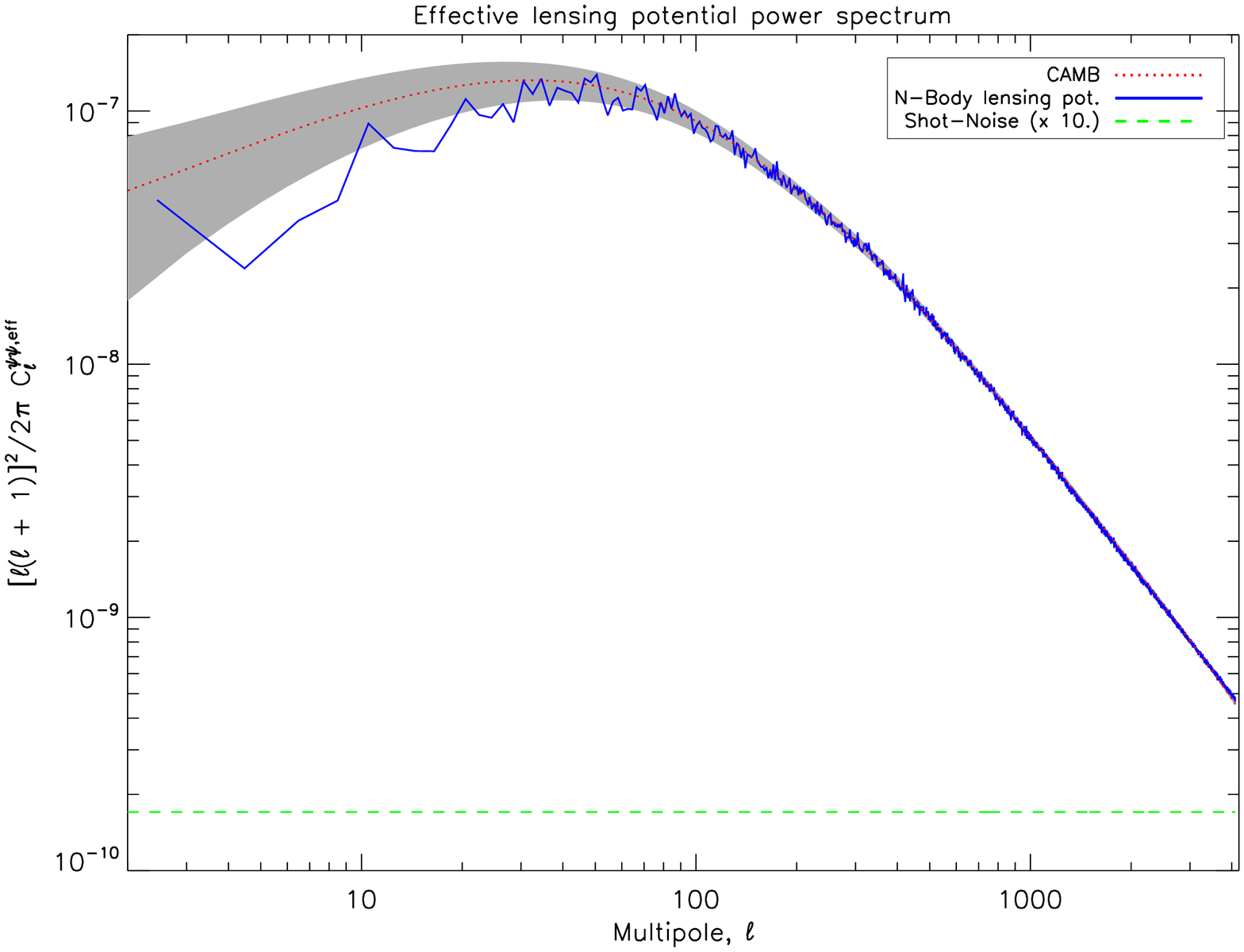}\\
\includegraphics[width=.38\textwidth,angle=90]{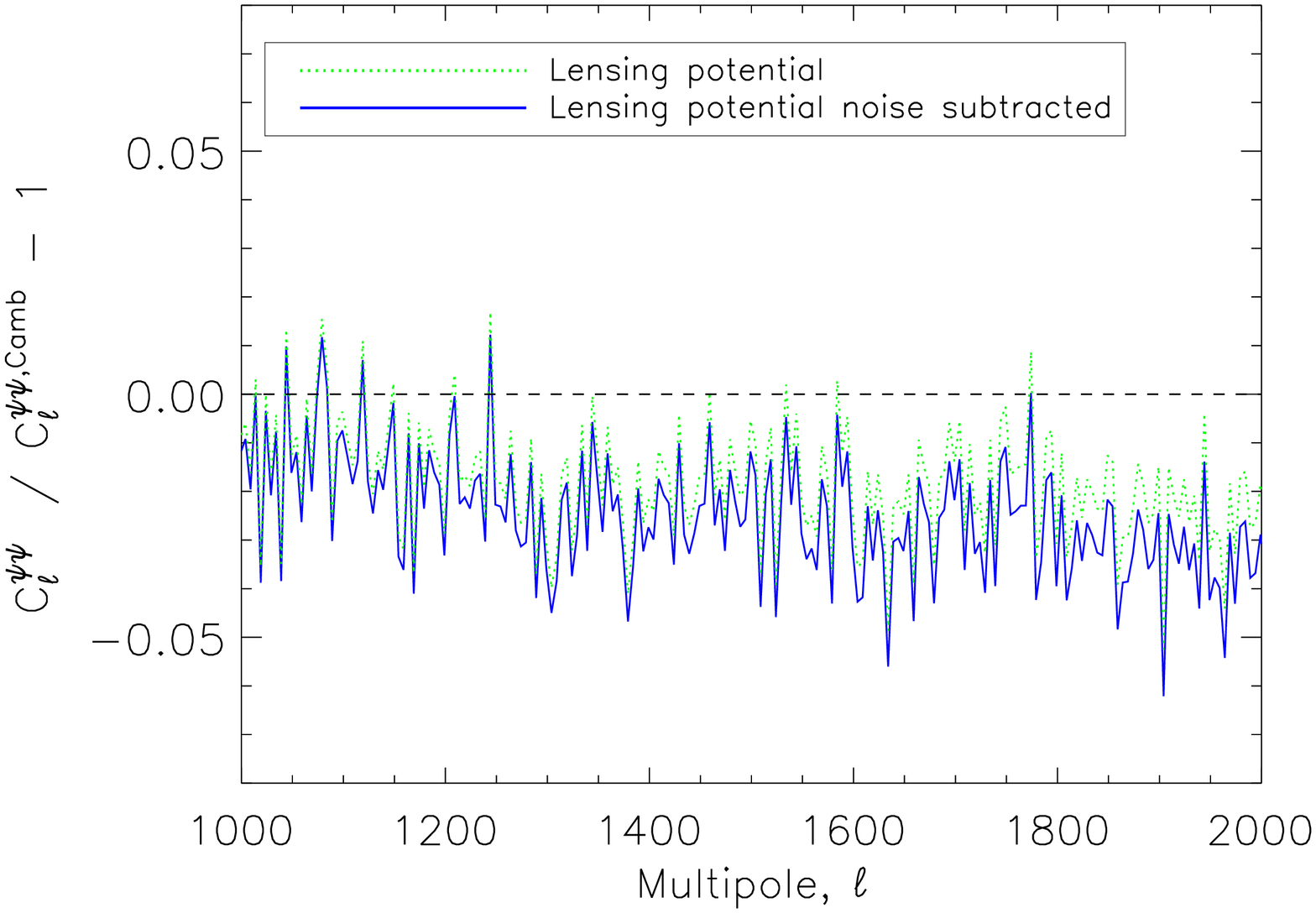}
\includegraphics[width=.38\textwidth,angle=90]{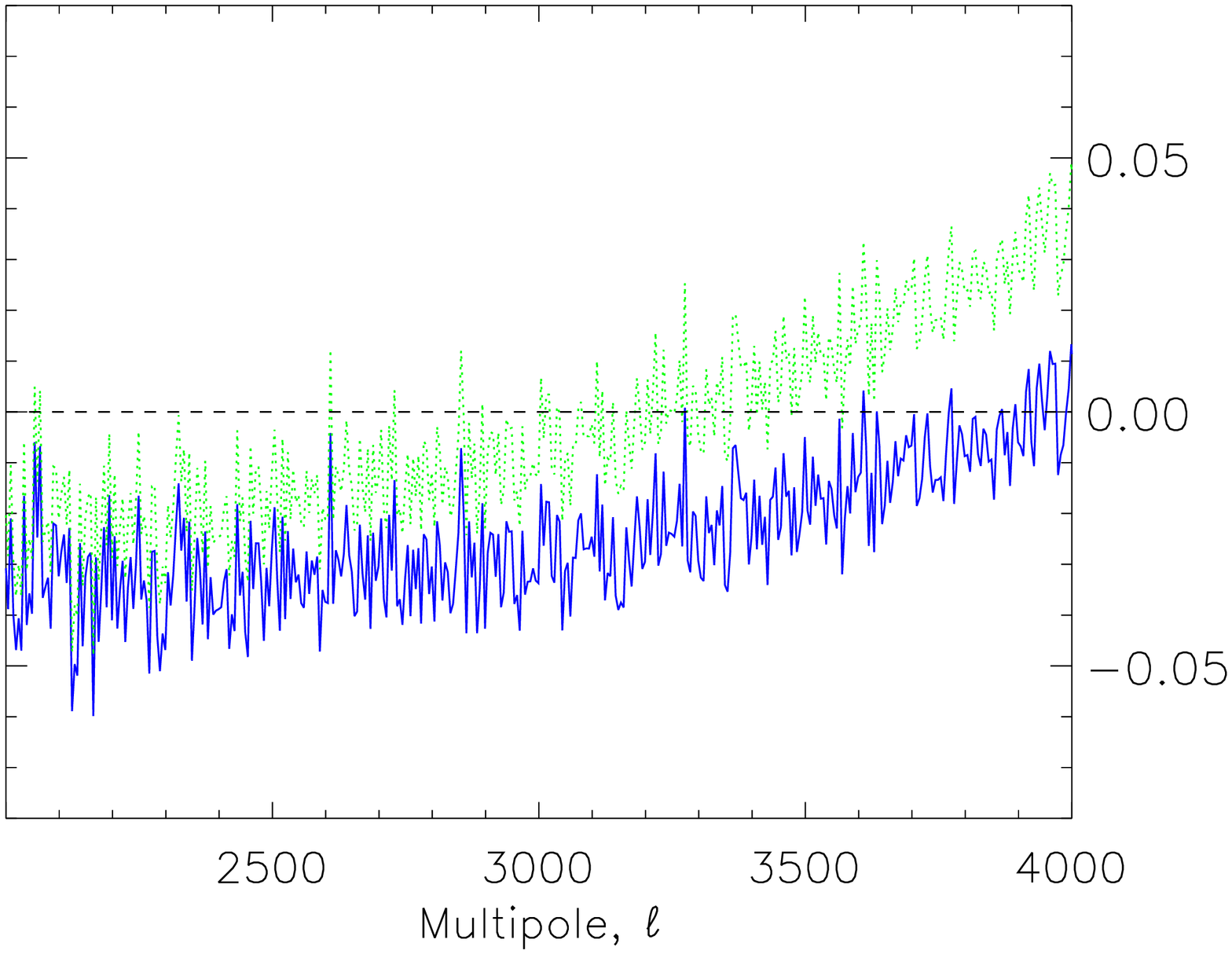}
\caption{Top panel: Integrated lensing potential $\psi^{eff}$ obtained
  in the Born
  approximation from the simulation (green) and the \lstinline!CAMB! result based on
  semi-analytical approximation of the non-linear evolution (red). The
  fractional difference between the two is displayed for some intermediate and high multipoles
  scales in the bottom panels. The results obtained after the shot-noise
  subtraction are displayed in blue solid lines. Note that the
  shot-noise spectrum in the top panel is multiplied by an arbitrary factor of 10
  to be seen clearly in the Figure.} 
\label{fig:phi_camb_subfig}
\end{figure}

Once surface density maps have been validated, we can move one step
forward and verify lensing quantities.
As described in Section~\ref{sec:discr}, the effective convergence plane
is computed by weighting the surface
mass density planes with appropriate
geometrical factors. We validated such convergence maps by
comparing the extracted power
spectra to the theoretical expectations based on semi-analytical
computations of the matter perturbation evolution as implemented
in the publicly available Boltzmann code \lstinline!CAMB!\footnote{\url{http://camb.info}}.
Adopting the Born approximation, we drew an ``effective'' convergence map, as described in
Eq.~(\ref{eq:born_kappa}), using the matter shells at different
redshifts. We then compute, in Limber approximation of
Eq.~(\ref{eq:clkappa}), the theoretical convergence angular power
spectrum, exploiting directly the 3D matter power spectrum computed
with \lstinline!CAMB!. The comparison between the simulated and the
analytical power spectra is shown in Figure \ref{fig:born_kappa}.
In both cases we perform the integration up to a specific
redshift $z^{*}$ to assess the validity of the maps at different
times. We observe that the measured spectra agree at high accuracy with the
theoretical predictions on a large interval of angular scales,
indicating the validity of our map-making procedure. As expected, the
lack of power at small multipoles $\ell\lesssim 50$ is due to the finite box size of the
N-Body simulation. 
A source of possible contamination of
the signal is the so-called {\it shot-noise}, due to the finite
particle density in the N-Body simulation. The shot-noise power spectrum can be computed analytically substituting the shot noise power spectrum $P^{Shot}(k, z)
= 1/\bar{n}_k$ in Eq.~(\ref{eq:clkappa_dis}),  where $\bar{n}_k = N_k
/\left(4\pi \eta^2_k \Delta \eta \right)$ and $N_k$ is the total
number of particles in the $k$-th shell, we obtain the shot-noise
contribution to the convergence:
\be
\label{eq:clkappa_noise}
C_{\ell}^{\kappa\kappa,{\rm Shot}} = \frac{9 H^4_0 \Omega^2_{m,0}}{4 c^4} \sum_k
  \Delta \eta \:
  \frac{1}{\bar{n}_k} \left( \frac{D_A(\eta_s - \eta_k)}{D_A(\eta_s)
    a_k} \right)^2.
\ee
For the N-body simulation used in this code the shot-noise is small at all redshifts given the high spatial resolution and high number of particles employed (see Figure~\ref{fig:born_kappa}).
\noindent
Figure~\ref{fig:phi_camb} shows a comparison of the partial
contributions to lensing potential
angular power spectrum computed at different redshifts
 with the corresponding analytical signal given by
Eqs.~(\ref{eq:clkappa_dis}) and (\ref{eq:kappa2lens})
in which we insert the 3D matter power spectrum extracted from
\lstinline!CAMB!. In this case, the
label $z$ refers to the redshift of the matter spherical map which
contributes to the 
lensing potential at that time. Each power spectrum
represents the ``real'' map given as input to \lstinline!LensPix! in
order to obtain the final CMB lensed maps in the multiple plane approach.
Here the Limber approximation is necessary 
to solve the Poisson equation using the
 transverse part of the Laplacian only, thus neglecting line-of-sight contributions as
previously discussed in Sec.~\ref{sec:discr}.
The agreement between simulated and
analytical $C^{\psi\psi}_{\ell}$ as a function of the redshift is
clearly observable from Figure~\ref{fig:phi_camb}.
The recovered signal is stable and coherent on
a whole range of multipoles. As discussed in the following, we assume
a very conservative choice for the map resolution and power cut-off
$\ell_{max}$ ($\textrm{\lstinline!NSIDE!}=\ell_{max} =
4096$). Therefore, we do
not expect this result to be affected by power aliasing
given that an \lstinline!HEALPix! grid with resolution set by \lstinline!NSIDE!
parameter should be able to properly sample modes up to $\ell\approx
2\times\textrm{\lstinline!NSIDE!}$.  
\noindent
An interesting and comprehensive way to see how the lensing process behaves at different scales is to look
at the integrated potential, as computed in Born approximation using
Eq.~(\ref{eq:born_phi_dis}). In Figure~\ref{fig:phi_camb_subfig} we show the angular power
spectrum for the effective lensing potential and its shot-noise
contribution, compared to the semi-analytical realizations by
\lstinline!CAMB!, where we fix the maximum
redshift of the integration, $z_{max}$, to be the same as the maximum
redshift used in our map-making procedure. Also in this case, we
find a very good agreement between the two methods, within the
1$\sigma$ uncertainty for the semi-analytical spectrum.
 As in Figures~\ref{fig:born_kappa} and ~\ref{fig:phi_camb} the spectrum
shows a lack of power due to the finite size of the simulation box for
$\ell < 50$. Note that the shot-noise contribution is negligible, as
we have multiplied it by a factor of 10 such that it could be compared with the
lensing potential signal.
In general, at intermediate scales our spectra show a small deficit of
power, within 3\% with respect to the HALOFIT prescription 
\cite{Takahashi12}, while at small scales, even after the shot-noise subtraction
(blue lines), the signal seems to increase towards $\ell\approx 3500$,
likely due to the underlying non-linear clustering underestimated by
the analytical models. Since the simulated spectra agree within percent level with the semi-analytical
realization of \lstinline!CAMB!, this means that our map-making
procedure traces with good accuracy the evolving matter distribution.

%% file: results.tex
 
\section{Results}

\subsection{Numerical effects of the pixel-based methods}\label{sect:numerical-pixel-based}
Pixel-based methods for CMB lensing, though in general very efficient, are subjected to several numerical problems.
The first one is related to the bandwidths of the lensed CMB fields
generated as a result of the calculation. 
Because the lensing effect happens before the intervention of any
instrumental response, the synthesis and analysis of relevant sky
signals in the pixel-based lensing methods (CMB source plane and
lensing potential map) require using a resolution sufficient to
support the signal up to the intrinsic bandwidth $\ell_{max}$ set by the user specific application and its required accuracy.
However, since mathematically the lensing effects act as a convolution in
the harmonic domain, the bandwidth of the resulting lensed field 
is broader than the one of the unlensed CMB and of the lensing
potential. Therefore, given the bandwidth used to synthesize the CMB source plane ($\ell_{max}^{CMB}$), i.e before undergoing any deflection, and the one used to solve the Poisson equation and to create the deflection field for a given shell ($\ell_{max}^{\psi}$), the
resulting lensed CMB will have an approximate band-limit of
$\ell_{max}^{CMB}+\ell_{max}^{\psi}$\footnote{We note that in general this bandlimit is only approximate and the resulting function is strictly not band-limited unlike the input source plane.}.
Consequently, the lensed map should
have its resolution appropriately chosen to eliminate potential power
aliasing effects on these angular scales \citep{Fabbian13}. We note
moreover that these aliasing effects are even more important
in the case of the ML approach because the bandwidth extension induced by
lensing happens each time the lensed CMB is propagated through a
single shell.\\*
The second challenge arises from the fact that
the displaced direction at each iteration ${\bf \hat{n}^{\prime}}{}^{(k)}$ does not correspond in general to the pixel
centers of any iso-latitudinal grid on the sphere. The values of the
CMB signal at those locations thus cannot be computed with the aid of fast
SHT algorithm and a more elaborated approach is needed.
In the context of pixel-based simulation methods, interpolation is the
most popular workaround of the need to directly calculate values of
the unlensed fields for every displaced directions 
The exact solution, which consist in a direct resummation of the
spherical harmonics at the displaced position, is in fact unfeasible even for moderate
resolutions \cite{Lewis05, Fabbian13}.
Any interpolation in this context, however, is not without its dangers
because interpolations tend to smooth the underlying signals and - as a
consequence - to hide aliasing effects in the lensed maps. For this
reason we chose the bandwidth of our signal
($\ell_{max}^{CMB}=\ell_{max}^{\psi}=4096$) and the resolution of our grid
(\lstinline!NSIDE!=4096) following the recipe provided in
\cite{Fabbian13} to minimize all of these effects simultaneously. This choice however limits the multipole range where the lensing signal can be reproduced with high reliability especially in the case of B-modes of polarization to $\ell\lesssim 2500$ (see Sec~\ref{sec:angular_power_spectrum}).

\subsection{Shot-noise contribution}
\label{sec:shotnoise}
\begin{figure}[htbp]
\centering
{\includegraphics[width=.55\textwidth,angle=90]{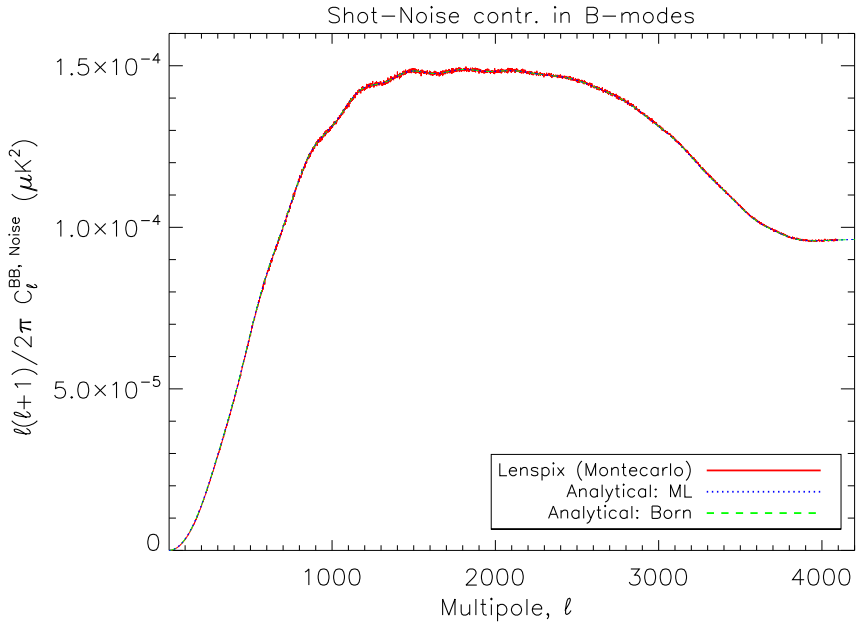}} 
\caption{Angular power spectrum for the lensed B-modes induced by the
  simulation shot-noise. The red, solid line is computed
  using 100 realizations of the algorithm in the Born approximation
  scenario. The green-dotted and the blue-dashed lines are evaluated
  with the analytical formula for - respectively - the Born and the
  multiple plane scenario. 
}
\label{fig:shot_noise_a} 
\end{figure}

In this section we estimate the impact of the intrinsic
discretisation of the N-Body simulation on the final lensed CMB power
spectra. Since we expect changes in the power spectrum on the order of few
percent, it is mandatory to be able to
control numerical artifacts with the same level of precision. 
For this study we use the analytical modelling of weak lensing in the harmonic
domain discussed in \citep{Hu2000}. This treats lensing as an
effective convolution in Fourier space between the unlensed CMB and
the lensing potential and it is based on a second order Taylor expansion
around the unperturbed photons direction. 
The formulae are accurate to better than the percent level on the
angular scales considered in this work, especially in the case of
B-modes, and allow to quantify more easily the impact of the choice of
the band-limit on the recovered result. In the specific case of
B-modes, the convolution reads:
\begin{equation}
\tilde{C}_{\tell_{B}}^{BB}=O(\tell_{B},C_{\ell}^{\psi\psi})\cdot C^{BB}_{\tell_B}
+ \frac{1}{2}\sum_{\ell_{E},\ell_{\psi}}\frac{| _2
F_{\tell_{B}\ell_{E}\ell_{\psi}}|^{2}}{(2\tell_{B}+1)}C_{\ell_{\psi}}^{\psi\psi}\left[C^{EE}_{\ell_{E}}(1-(-1)^{L}) +C^{BB}_{\ell_{E}}(1+(-1)^{L})\right] , 
\label{eq:analyticalb}
\end{equation}
where we denote with tilde a lensed quantity and
$L=\tell_B+\ell_{E}+\ell_{\psi}$. We refer the reader to
\citep{Fabbian13} for a detailed discussion of the properties of the
convolution kernels $F_{\tell_{B}\ell_{E}\ell_{\psi}}$ and to the
$O(\tell_{B},C_{\ell}^{\psi\psi})$ factor. Similar expression can also
be derived for the TT, TE and EE power spectra \citep{Hu2000}.
Since this formalism does not make any assumption on the explicit form
nor the origin of $C_{\ell}^{\psi\psi}$, we can plug in
Eq.~(\ref{eq:analyticalb}) the shot-noise power spectrum instead of the
lensing potential extracted from the N-body simulation, to estimate the
fraction of the recovered signal generated by the limited resolution
of our simulated data.
For this purpose we truncated the sum of Eq.~(\ref{eq:analyticalb}) to
the same band-limit value used in the lensing simulation,
i.e. $\ell^{E}_{max}=\ell^{\psi}_{max}=4096$. \\*
We first evaluate the shot-noise contribution to the $\psi^{eff}$
lensing potential starting from Eq.~(\ref{eq:clkappa_noise}) and
assuming the average number density to be the one of the
$\psi^{eff}$ field, $\bar{n}^{eff}_{k}$. This is then used as an input
for the analytical formulae of Eq.~(\ref{eq:analyticalb}), assuming the
primordial B-modes $C_{\ell}^{BB}$ to be zero as it is the case for
the unlensed CMB realizations used in the following. To validate the
analytical shot-noise predictions we also produce 100 Monte Carlo
realizations of shot-noise for the effective lensing
potential. We use those maps to extract a deflection field which is
then given as input to \lstinline!LensPix! to lens a random
realization of the CMB signal. The final average of all the power
spectra of these set of lensed CMB maps contains thus only the
lensing effect due to the shot-noise acting on primordial
anisotropies.\\*
To evaluate the shot-noise contribution to the ML method we compute 
Eq.~(\ref{eq:clkappa_noise}) for each $k$-shell and then apply the analytical convolution 
iteratively assuming as input CMB spectrum for the $k$-th shell the lensed one
emerging from lensing of the previous $(k-1)$-th iteration. \\*
As discussed in the
following section, if we assume a power cut-off for the incoming CMB
and the lensing potential to be $\ell_{max}^{CMB}$ and
$\ell_{max}^{\psi}$ respectively, the lensed CMB after each deflection shows
power up to a multipole $\tell \approx  \ell_{max}^{CMB}+\ell_{max}^{\psi}$,
due to the properties of the lensing convolution kernels in the harmonic domain
\citep{Fabbian13}. The evaluation of the lensing kernels
requires in fact a computationally-heavy summation of Wigner-$3j$
symbols up to high multipoles. We therefore have assumed that for the
iteration $k>0$ the incoming CMB has power at most up to
$\ell=8192$. This additional cutoff is high enough not to affect
significantly the results on the scale considered in this work. The
analytical formulae were validated also in this case with Monte
Carlo simulation where for each shell the noise realizations were
generated starting from the shot-noise power spectrum of the single
shell.\\*
In Figure~\ref{fig:shot_noise_a} we show the results obtained from
both these methods, for the B-mode power spectrum, which is the most sensitive to
the details of the lensing potential being entirely lens-induced in
our case (no primordial tensor modes). The TT and EE spectra are 
conversely quite insensitive to the
the shot-noise which impacts the results at the sub $0.1\%$ level
(see Figure~\ref{fig:tt_ee_lensed}).
Both the analytical and Monte Carlo estimates of the shot-noise
contribution in the Born approximation agree extremely well
at all angular scales. The shot-noise contribution in the ML approach
is comparable to the effective case at $\ell\lesssim 2000$ though the difference is less than $0.5\%$. 

\subsection{Angular power spectrum}
\label{sec:angular_power_spectrum}
\begin{figure}
\centering
{\includegraphics[width=.55\textwidth,angle=90]{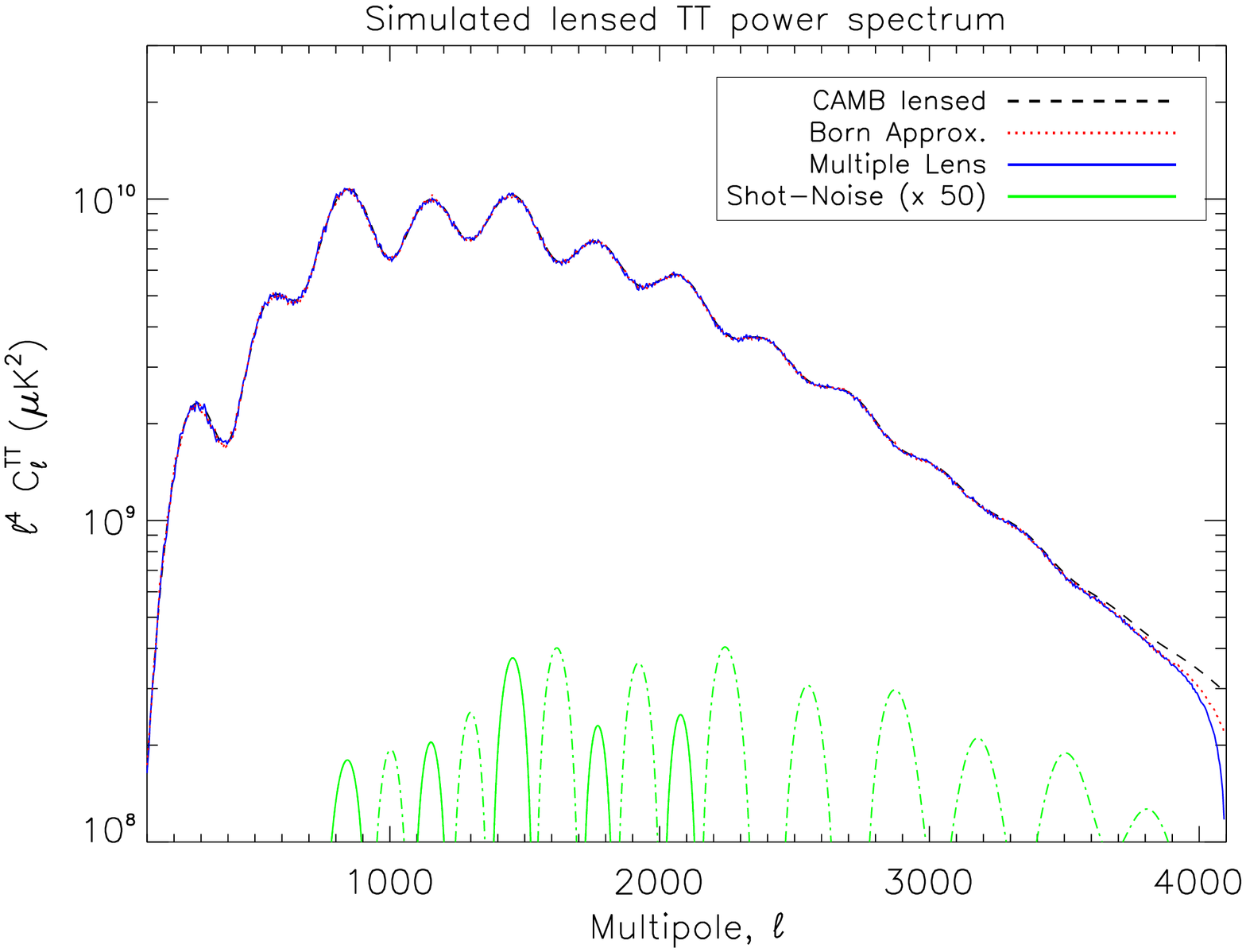}} \\
{\includegraphics[width=.55\textwidth,angle=90]{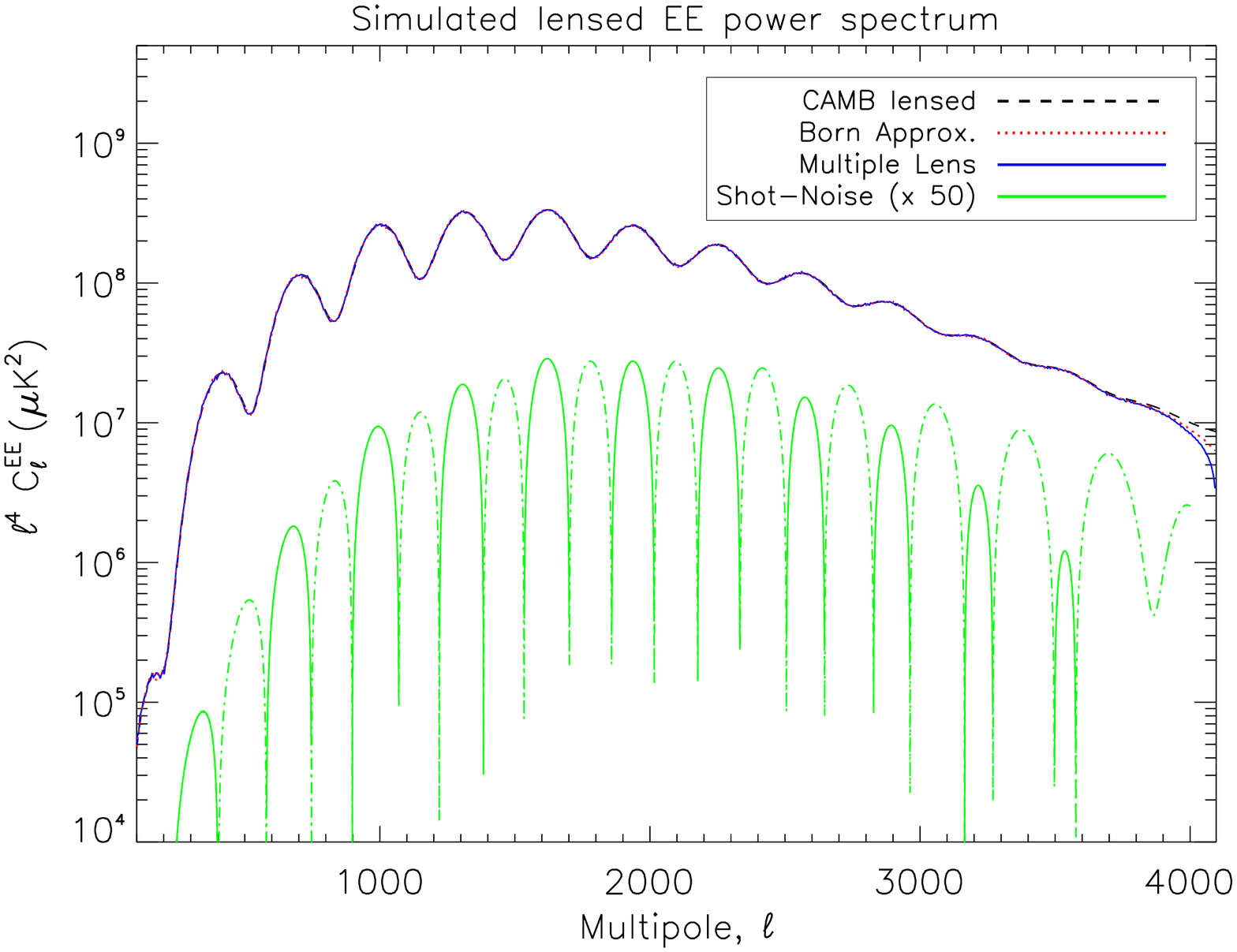}} 
\caption{
Angular power spectrum of the (lensed) total intensity T (top panel) and 
the E-mode polarization of the CMB (bottom panel). Black dashed lines are \lstinline!CAMB! realization of a
lensed spectra. Red dotted
line uses the lensing field as in the Born
approximation, while for the blue solid line the CMB is lensed through
multiple planes. Green lines in both panels show the shot-noise
contribution to the lensed TT and EE spectra; green,
dot-dashed lines represent the absolute value of this
contribution. Note that the shot-noise
power spectra is multiplied by a factor of 50 for visualization
purposes. }
\label{fig:tt_ee_lensed} 
\end{figure}

\begin{figure}
\centering
\includegraphics[scale=0.6,angle=90]{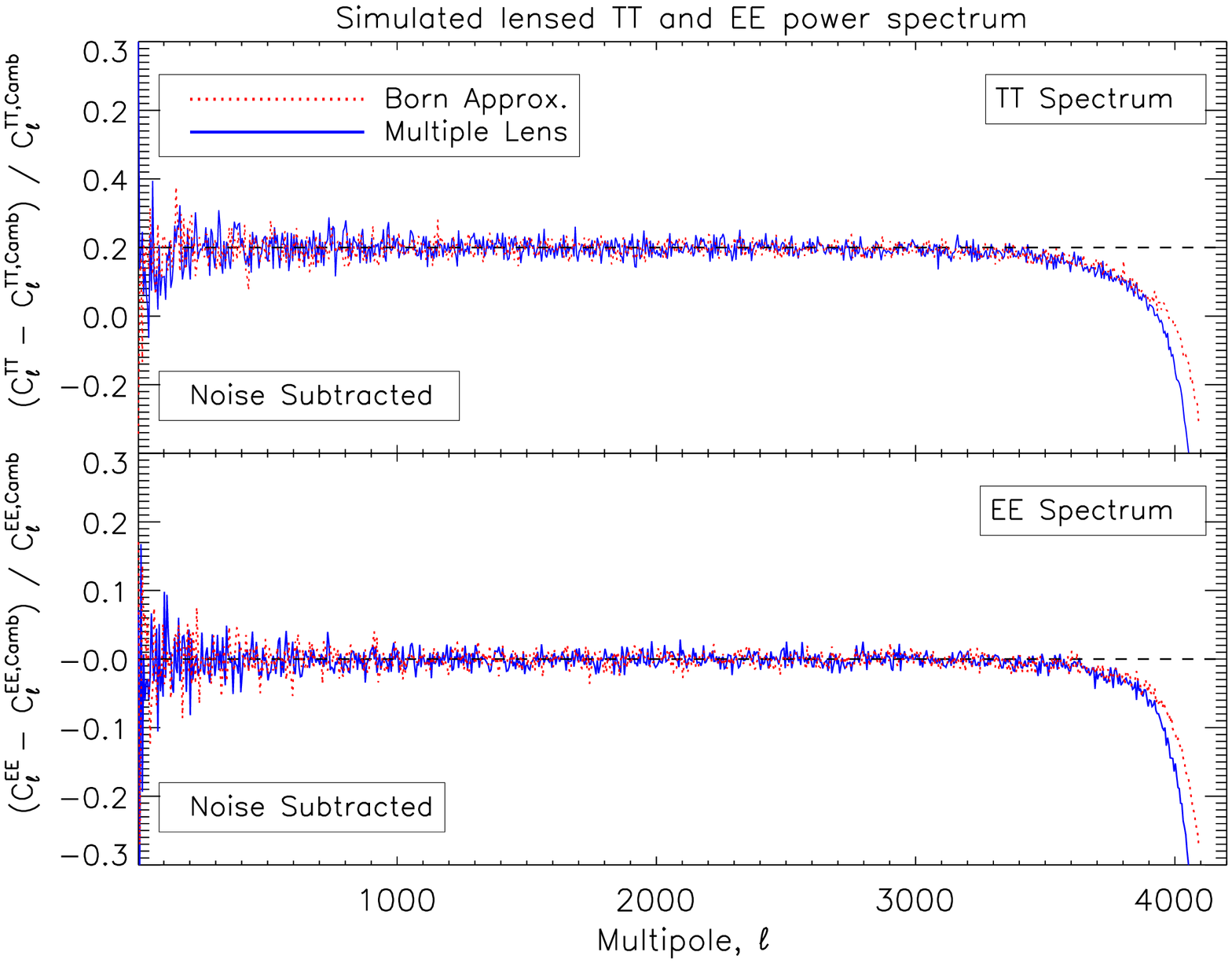}
\caption{
Fractional difference for the angular power spectrum of the temperature
(top panel) and E-mode polarization (bottom panel) with respect to
\lstinline!CAMB!. Red dotted
lines are obtained in the Born
approximation, while blue solid in the multiple lens plane
approach. The shot-noise has been subtracted in both cases.
}
\label{fig:tt_ee_lensed_ratio}
\end{figure}

Similarly to the case of the lensing potential extraction,
we now take into consideration two different approaches also 
for the evaluation of the angular power spectrum of the lensed CMB. 
The first set of primary CMB maps are lensed in the Born
approximation,  while the second set by mean of the ML approach.
In Figure~\ref{fig:tt_ee_lensed} we show the comparison between the expected CMB lensed temperature and the E-modes of polarization power spectra, ($C^{TT}_{\ell}$ and
$C^{EE}_{\ell}$), estimated using semi-analytical halo
mass function implemented in \lstinline!CAMB! \cite{Smith03, Takahashi12}, and the spectra
extracted from our lensed CMB maps. For both these cases the simulated power spectra
follows precisely the \lstinline!CAMB! signal. In particular, the
shot-noise-induced contribution (evaluated following the recipe of the
Section~\ref{sec:shotnoise}) for these two observables is negligible
given that the effect of lensing per-se is already minor. Thus,
changes introduced at percent variation in the lensing potential are
further mitigated and hidden in the numerical noise. After having
subtracted the shot-noise induced lensing contribution, the fractional
difference between \lstinline!CAMB! and the N-Body lensed spectra
shows no significant bias up to $\ell \approx 3000$ where we start
seeing effects due to the choice of $\ell_{max}^{CMB}$. The latter is
not high enough to properly resolve power on those scales with
high-accuracy. The difference between the results obtained with the
Born and ML method is negligible and important only towards scales
$\ell\approx\ell_{max}$ (see Figure~\ref{fig:tt_ee_lensed_ratio}). The
abrupt decrease in power observed on those scales for the ML approach
with respect to the Born approximation is due to the effect of polynomial
interpolation. As the latter tends to smooth the underlying signal,
the consecutive application effectively removes more power on small
angular scale with respect to the Born approach, for which the
interpolation is performed only once.


\begin{figure}
\centering
\includegraphics[scale=0.7,angle=90]{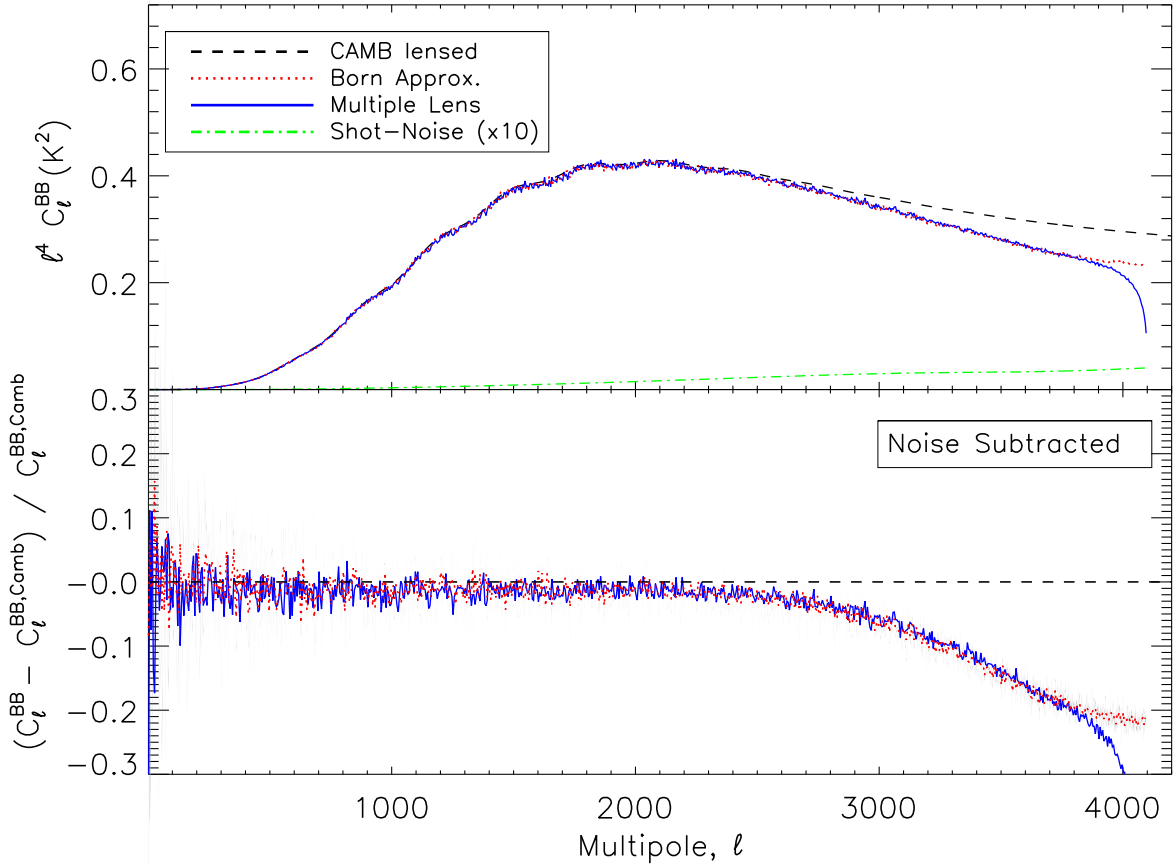}
\caption{Angular power spectrum for the B-mode polarization of CMB. Black dashed lines are
  \lstinline!CAMB! realization of a lensed spectra. Red dotted lines
  are obtained in the Born
approximation, while blue solid lines in the multiple lens
planes approach. The green dot-dashed line is
a lensed spectrum produced from a shot-noise-only lensing map. Note that the shot-noise
power spectra is multiplied by a factor of 10 for visualization
purposes. 
In the bottom panel, it is shown the fractional difference
with the reference \lstinline!CAMB! spectrum. In this panel, the noise power spectrum
  has been subtracted from the original signal.
}
\label{fig:bb_lensed_noise}
\end{figure}

The situation however is different for the B-modes of polarization, as
shown in  Figure~\ref{fig:bb_lensed_noise}. This signal is 
entirely caused by lensing as we have set the primordial tensor
modes to zero. Its behaviour is therefore a clear imprint of how the LSS
process the primary CMB field and thus we expect this observable to reflect more directly the features observed in the lensing potential. As expected from the analysis of the lensing
potential in Section~\ref{sec:lensing_pot}, the BB spectrum shows
a lack of of power at percent level with respect to \lstinline!CAMB! spectra,
in agreement with the matter power spectrum of the N-Body
simulation, though this effect is partially compensated by the increase in power at small scales in the lensing potential. This feature is not observable in the lensed T or E field, where power coming from primordial
anisotropies is dominant over the lensing-induced one. Moreover, 
while negligible in the TT and EE cases, we found the shot-noise contribution to be
important at the percent level for the BB power spectrum at small scales. This is
expected given that B-modes are very sensitive to non-linear power,
which is affected by shot-noise for $\ell >> 2000$ (see bottom panel
of Figure~\ref{fig:phi_camb_subfig}). The lack of power due to the
choice of $\ell_{max}^{CMB}$ starts to be important on angular scales
larger than the ones affected in T and E-modes power spectra. This can
be explained considering that at those scales a non-negligible
fraction of the contribution to the BB power spectrum starts to come
from progressively higher multipoles of both E and $\psi$. At $\ell_B
\sim 4000$, for instance, a 25\% contribution to the power in the
B-modes comes from scales in the E and $\psi$ fields at $\ell >
\ell_{max}=4096$ \citep{Fabbian13}. Since our algorithm is
band-limited to this $\ell_{max}$,  cutting power for those high
$\ell$, produce a loss of about 25\% in the BB power spectrum at that
particular multipole (as shown in Figure~\ref{fig:bb_lensed_noise}). 

\begin{figure}
\centering
\includegraphics[scale=0.6,angle=+90]{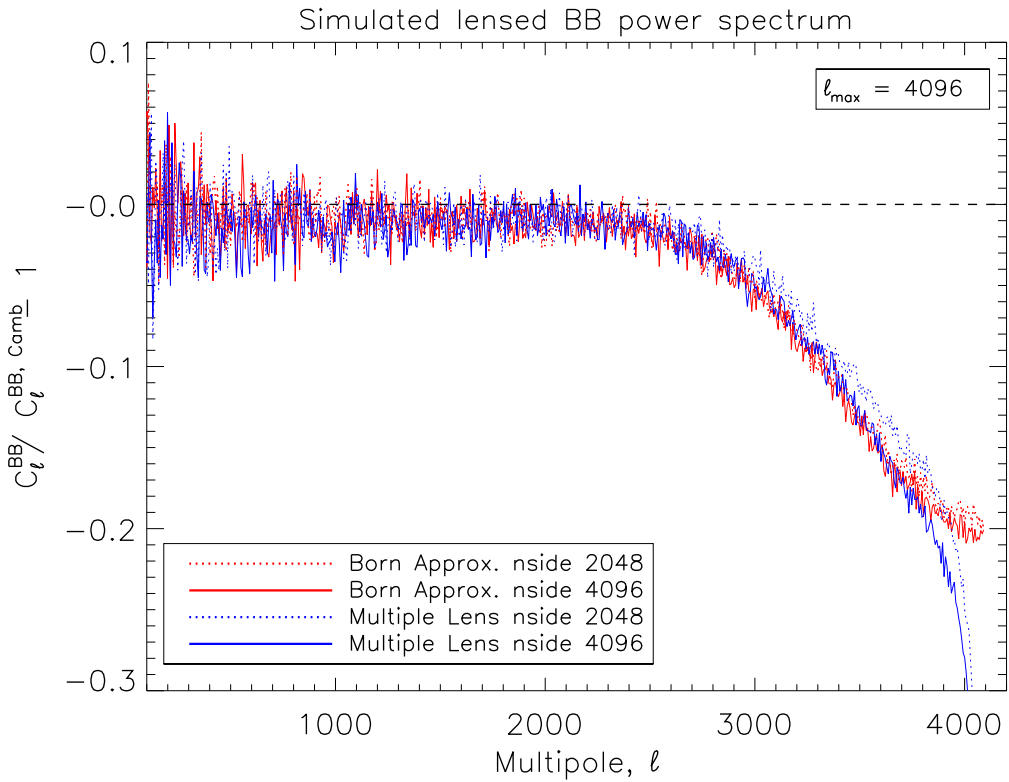}
\caption{
Fractional difference of
angular power spectrum of the BB power spectrum with respect to the reference \lstinline!CAMB! spectrum. Dotted lines refers to the map at low resolution (\lstinline!NSIDE!=2048), while solid lines
to the map with \lstinline!NSIDE!=4096. Red lines plot the effective,
Born approximation case, blue lines are connected to the multiple
plane approach. Note that for this
comparison, the noise power spectrum
has not been subtracted from the original signal.
}
\label{fig:resolution}
\end{figure}

%
%
%
As argued in Section~\ref{sect:numerical-pixel-based}, one of the major numerical problem affecting the simulation of CMB B-modes is the power aliasing due to bandwidth extension induced by lensing.  In Figure~\ref{fig:resolution} we show the impact of this effect as a function of the map
resolution on the B-modes power spectrum recovery. For this tests we
extracted the lensing potential maps for both Born and ML approach
using two different HEALPix grid at $\textrm{\lstinline!NSIDE!}=2048,
4096$ and refer to these two setup as the low and high resolution case
respectively. We then synthesized on the same grid the CMB source
plane assuming the same bandlimit $\ell^{CMB}_{max}= 4096$, as done
for the results discussed above, and propagate it through the lensing
plane(s). As shown in Figure~\ref{fig:resolution}, the Born
approximation method is quite insensitive to the choice of
\lstinline!NSIDE! because the polynomial interpolation is effective in
removing most of the aliasing. However, for the ML scenario the
situation is worse as the aliasing generated by multiple deflection
can add up, becoming progressively more important.
This can then lead to a misinterpretation of the result obtained using
the ML, which seems to be significantly different from the once
obtained in the Born approximation. The fact that this difference
vanishes in the high-resolution case is a demonstration of the high
level of control of numerical effects which needs to be achieved for
this kind of algorithms. Even though these effects were limited in the
setup considered here, we expect those to become more important when
targeting accurate lensing simulations on scales $\ell>>2000$.
\\

Finally, we compare the differences in the angular power spectra between the Born approximation and the
multiple planes approach. First we define the quantity $O^{X}_{\ell}$
as the difference between the angular power spectra extracted with the
multiple lens approach and the one computed in the Born approximation, 
\be
O^{X}_{\ell} = C^{X, {\rm ML}}_{\ell} - C^{X, {\rm Born}}_{\ell}, 
\ee
where X = TT, EE, BB. Its uncertainty is given by the cosmic variance affecting both
spectra, or
\be
\sigma_{O^X_{\ell}} = \sqrt{\frac{2}{2\ell+1}\left( | C^{X, {\rm
      ML}}_{\ell}|^2 + |C^{X, {\rm Born}}_{\ell}|^2 \right)}.
\ee
Starting from these quantities we can define a reduced
chi-square $\tilde{\chi}^2$ statistics
\be
\tilde{\chi}^2 = \frac{1}{\ell_{max}-1} \sum^{\ell_{max}}_{\ell = 2}
\frac{O^2_{\ell}}{\sigma^2_{O_{\ell}}}, 
\label{eq:chi2-error}
\ee
to assess whether the two methods are inconsistent.
Since we expect $O^X_{\ell}/ \sigma^{2}_{{O_{\ell}}}$ to be a Gaussian
random variable,
we can we also perform a Kolmogorov-Smirnov (KS) to test
whether this hypothesis is verified or systematic differences exists
between the two methods. In defining both theses tests and the sample
variance of Eq.~(\ref{eq:chi2-error}), we assumed that the covariance
of the lensed power spectra is Gaussian. This assumption neglects the
fact that lensing introduces non-Gaussian correlations between
different modes \cite{benoit-levy12, KSmith04}, but this effect is
mainly important for B-modes, for which the gaussianity assumption
underestimates the sample variance.
 
In Table~\ref{table:tab1} we report the results of both those tests
expressed as the significance level probability.
In both cases we find that the power spectra obtained in the Born
approximation and with the ML method are statistically consistent. A
further possible test to compare the two methods would be to
reconstruct the effective integrated matter density from the simulated
lensed CMB maps as done in \cite{antolini}, but 
we leave this option to future work. 
\begin{table}
\centering
\begin{tabular}{lcc}
\toprule
$C^{X}_{\ell}$ & Significance $P_{KS}$ & Significance $P_{\tilde{\chi}^2}$\\
\midrule
TT & 0.47 & 0.70\\
EE & 0.19 & 0.51\\
BB & 0.21 & 0.19\\
\bottomrule
\end{tabular}
\caption{\label{table:tab1} Results of statistical tests on difference
  between lensed CMB angular power spectra in the Born approximation
  and multiple lens planes approach. The significance level
  probability for the null hypothesis using a Kolmogorov-Smirnov test
  ($P_{KS}$) and a reduced chi-square $\tilde{\chi}^2$ statistics
  ($P_{\tilde{\chi}^2})$ show no difference between the power
  spectra computed with the two methods on a statistical level.
}
\end{table}


%% file: conclusions.tex

\section{Discussion and Conclusions}
\label{sec:concl}

We have developed and tested a new algorithm to study the gravitational lensing of the
CMB on the full sky. Starting from snapshots of an N-Body
simulation, we reconstructed the whole
lightcone around the observer between $z\in [0,10]$ overcoming the
finite size of N-Body box through the box stacking technique developed
in \cite{Carbone08}. We sliced the
lightcone into 25 different spherical maps onto which the matter
distribution was projected. The spherical shells were then used as
source planes to lens the incoming CMB photons either adopting an
effective method based on the Born approximation or using a
multiple-lens ray-tracing approach. This is, to our knowledge, the
first attempt to apply this kind of algorithm to the CMB polarization. 
%
For this reason, we performed a detailed analysis of the numerical effects involved in the ML method coming both from the N-Body simulation and from the ray-tracing procedure itself. 
The Born approximation, which has been widely tested in the literature, was used as benchmark to
highlight the multiple-lens range of validity and to asses its virtues in
reproducing non-linearities from the N-Body simulation at small scales. 
In particular, the projection of the N-Body matter distribution onto
concentric spherical maps allows to compress all the interesting
information from the N-Body simulation 
into a more manageable lightcone, mimicking a realistic distribution of large scale structure as
observed by present and future large galaxy surveys.\\* 
%
%
We validated the lensing planes reconstruction both on the map level
and on the statistical level using the 2-point correlation function in
the harmonic domain evaluated for all the spherical maps constructed
across 
the past lightcone. We found the latter to be reproduced fairly well
by semi-analytical approximations to the non-linear evolution
implemented in widely used Boltzmann codes, though deviations at the
percent level were clearly observed.
We also analysed the final, lensed CMB anisotropies in both temperature and
polarization for the effective as well as the multiple
plane approach, paying a particular attention to the B-modes of polarization. 
These are in fact the most sensitive quantities both to the overall lensing
process and to the numerical effects. In the latter case, we discussed
in detail how to minimize their impact.
We found however that these numerical effects are usually negligible for the temperature and E-modes polarization field and important only for B-modes. The B-modes signal was found to be lower than the one computed using the semi-analytical, following the general trend observed in the extracted lensing potential power spectrum.\\*
%
Finally, we have extended the control of the validity of the Born approximation to the limiting 
resolution of the present setup of our simulations. Our results indicate that, when checking the 
angular power spectra of lensing observables, including CMB lensed fields, the latter approximation 
describes well the ray-tracing performed by the ML approach. However we expect the latter to perform 
better for studies aiming at investigating the statistics of the signal at smaller angular scales, 
or in presence of distortion from isolated, sharp structures. \\*
For what concerns the behaviour of the signal at large
multipoles $l>3000$, in a recent paper \cite{Hagstotz14} the authors model corrections to the Born
approximation by using perturbation theory applied to the lensing magnification matrix. 
This kind of analysis was first presented in \cite{CoorayHu02} and \cite{ShapiroCooray06}, who 
computed the second-order corrections to the angular power spectra of 
the lensing observables (convergence, shear and rotation). 
By adopting the Peacocks \& Dodds  matter 
power spectrum \cite{PeacockDodds96}, they concluded that these corrections are not relevant for the galaxy 
weak-lensing case being two orders 
of magnitude lower than the first-order contribution. On the other 
hand, using the matter
 power spectra extracted from \lstinline!CAMB!, \cite{Hagstotz14} applies 
the same analytical framework of \cite{CoorayHu02} and
\cite{ShapiroCooray06} to the CMB-lensing 
case, and finds a large excess in the B-modes power spectrum with respect to 
the first-order contribution. For a cross-check of these results, following the same analytical setup, we have 
independently computed second order corrections to the CMB convergence power 
spectrum, using the \lstinline!CAMB! matter power spectra as input. In agreement with 
\cite{Hagstotz14}, we find that these corrections seem to affect the CMB
 lensing potential at very small scales and, consequently, the B-mode power
 spectrum at all the multipoles. We have also applied the same analytical second order computations to the galaxy weak-lensing
 case with sources at $z_s=1,2$, using again \lstinline!CAMB! matter power spectra 
as input. In this case, we find in the convergence power spectrum an
 excess of power at $\ell \approx 2000$ of the order of $10\%$ with 
respect to first order contributions, largely 
in tension with the numerical
analyses of multiple-lens ray-tracing from N-body simulations present in the literature \cite{Hilbert09,Becker12}, which conversely find no
evidence of important differences with respect to the Born 
approximation for $z_s=1,2$. \\*
Given this tension, it is necessary to investigate in more detail the 
validity of the analytical corrections. To this purpose, we are developing an 
improved numerical setup to properly simulate both the CMB lensing 
per se and the LSS evolution below the smallest angular scales
 considered in this work.  
We plan to address this issue in a future paper including also an 
extension of our formalism to propagate the whole lensing
 magnification matrix, in order to trace more accurately all second-order
corrections for all the lensing observables. 

%
The spherical map matter projection, in both the Born
approximation and ML implementations, can be particularly useful in
cross-correlation
studies between the CMB with other tracers of mass 
and foreground sources, for characterizing the simulation of mock
catalogues of observables built from N-Body simulations. The
tomography of LSS, which is an intrinsic feature of the two lensing
approaches analysed in this work,
can be exploited to investigate different cosmological scenarios,
looking at the effects of different DE models on small scales as well
on the whole evolution of the matter in Universe. These feature will
be of great importance for upcoming projects such as the Euclid
satellite that can fully exploit the capabilities of cross-correlation
as cosmological probe.